\documentstyle[11pt,epsfig]{article}
\begin{document}

\begin{flushright}
{hep-ph/0310247}\\

\end{flushright}
\vspace{1cm}
\begin{center}
{\Large \bf The anomalous Higgs-top couplings in the MSSM}\\
\vspace{0.5cm}
\end{center}
\vspace{1cm}
\begin{center}
{Tai-Fu Feng$^{a}$\hspace{0.5cm}Xue-Qian
Li$^{b,c}$\hspace{0.5cm}Jukka Maalampi$^{a}$\\} \vspace{.5cm}

{$^a$Department of Physics, 40014 University of Jyv\"{a}skyl\"{a} ,
 Finland}\\
{$^b$ CCAST (World Laboratory), P.O. Box 8730, Beijing 100080,
China}\\
{$^c$Department of Physics, Nankai
University, Tianjin 300070,  China}\\
\vspace{.5cm}

\end{center}
\hspace{3in}

\begin{center}
\begin{minipage}{11cm}
{\large\bf Abstract}

{\small The anomalous couplings of the top quark and the Higgs
boson have been studied in an effective theory deduced from the
minimal supersymmetric extension of the standard model (MSSM) as
the heavy fields are integrated out. Constraints on the parameters
of the model from the experimental data of
$R_b={\Gamma(Z\rightarrow b\bar{b})/\Gamma(Z\rightarrow hadrons)}$
are obtained.}
\end{minipage}
\end{center}

\vspace{4mm}
\begin{center}
\end{center}

\section{Introduction}
\indent\indent The standard model (SM) has been very successfull phenomenologically,
but nevertheless it should be considered just as an
effective theory valid for physics at the electroweak (EW) scale. In higher-energy
regimes new physics beyond the SM must exist. Irrespective to what this new
physics might be, it should be able to give a satisfactory answer to
the most fundamental open question of the electroweak physics, that is,
it must explain the origin of  the electroweak gauge symmetry
breaking  \cite{Dawson,Carena}. In the SM this  is arranged through
the spontaneous symmetry breaking mechanism by introducing a doublet
of scalars with a nonzero vacuum expectation value (VEV).
This mechanism, despite its simplisity and economy,  has
 well known problems, which has enforced theorists
and experimenters to look for new physics beyond the SM.
Among the possible ways of extending the SM, supersymmetry is considered
as a particularly attractive one. The minimal supersymmetry extension of
the standard model (MSSM) provides an appealing solution to the gauge
hierarchy problem by guaranteeing the perturbative stability of the theory
from the electroweak scale to the Planck scale.

The MSSM contains two complex Higgs doublets, denoted by
$H_{_u},\; H_{_d}$ and assigned with opposite hypercharges
$Y_{_B}(H_{_u})=-Y_{_B}(H_{_d})=1$. There are altogether four
neutral scalar degrees of freedom, three of which correspond to
physical scalar fields. In the case where $CP$ is conserved one
can define two $CP$-even neutral Higgs fields, $H,\;h$, and one
$CP$-odd neutral Higgs field, $A$. The present experimental bounds
on the Higgs boson masses set strong restrictions on the parameter
space of the $CP$-conserving MSSM  \cite{higgs2}. Radiative
corrections to the lightest $CP$-even Higgs boson mass have been
computed by using the renormalization group equation (RGE) method
and diagram technique  \cite{higgsm}, and the resulting upper
bound is $135\;{\rm GeV}$, which is not much above  the present
experimental lower bound of $95 {\rm GeV}$ ($95\%$ CL).

The possibility of $CP$ violation  makes the situation drastically different. There
are three main sources of the $CP$ violation  in the
MSSM Lagrangian. The first one is the well known $\mu$ parameter of the
superpotential, which is in general complex. The second source is constituted
by the soft mass terms of the $SU(3)\times SU(2)\times
U(1)$ gauginos. The third source are the phases of the soft
supersymmetry-breaking mass terms of scalar fermions and of
the soft trilinear couplings, which are presented by the matrices
${\bf m}_{_{Q,U,D,L,R}}^2$ and ${\bf A}_{_{U,D,E}}$, respectively.
Actually,  only the off-diagonal elements of the soft mass matrices
can be complex due to the hermiticity of these matrices.
The matrices ${\bf A}_{_{U,D,E}}$ in contrast can have  complex phases
also in their diagonal elements  \cite{Brhlik}. Not all the phases of
these soft SUSY breaking parameters  are physical and
lead to the violation of $CP$ parity.  The physical $CP$ phases
are restricted by experimental observations, the most rigorous
constraints originating from the measurements of the
electron and neutron electric dipole moments (edm).
The present upper limits for these edms are
$d_{_e}< 4.3\times 10^{-27} {\rm e}\cdot {\rm cm}$  \cite{Commins} and $d_{_n}<
6.5\times 10^{-26} {\rm e}\cdot {\rm cm}$ \cite{Harris}, respectively.
Also the edm of $H_{_g}^{199}$ is quite accurately measured,
the present upper limit being $d_{_{H_{g}^{199}}} < 9.\times
10^{-28} {\rm e}\cdot {\rm cm}$  \cite{Lamoreaux}.

It has been demonstrated that the MSSM can be consistent with
these constraints in some regions of the parameter space when
suitable cancellations between different contributions occur
\cite{Ibrahim1} or when $CP$ violation effects are associated with
the third generation of squarks only \cite{Chang}. The mixing of
neutral Higgs bosons in the latter scenario is analyzed in
\cite{Pilaftsis1,Pilaftsis2,Pilaftsis3,Pilaftsis4}. It is found
that the $CP$-violating phases and  large Yukawa couplings of the
third generation fermions can lead to large mixings among the
neutral Higgs bosons as a consequence of radiative effects. These
mixings can drastically change the couplings between the neutral
Higgs bosons and quarks and between the neutral Higgs bosons and
gauge bosons, as well as the self-couplings of the Higgs fields.
One consequence of this is that the experimental lower bound on
the lightest neutral Higgs mass is relaxed to $60\; {\rm GeV}$,
while the predicted upper limit for the lightest Higgs boson mass
remains about $135\; {\rm GeV}$.

If the new physics scale is much higher than the EW scale, one would have
at the EW scale a great number of higher-dimensional operators
${\cal O}_i$ ($dim({\cal O}_i)>4$) induced by the beyond-the-SM
physics \cite{Burgess,Leung, Buchmuller, Hagiwara}. The resulting
effective Lagrangian is of the general form
\begin{eqnarray}
&&{\cal L}_{_{eff}}={\cal L}_0+{1\over \mu_{_{NP}}^2}\sum\limits_i
C_i{\cal O}_i+{\cal O}({1\over \mu_{_{NP}}^4})\;.
\label{elag}
\end{eqnarray}
Here ${\cal L}_0$ is the SM Lagrangian, $C_i$ are Wilson coefficients,
and $\mu_{_{NP}}$ the  energy scale of new physics. The
Wilson coefficients are in general dependent on the new energy scale,
but in addition to this all the higher-dimensional operators in
${\cal L}_{_{eff}}$ have a common suppression factor $1/\mu_{NP}^2$.

In this paper we shall study anomalous couplings (to use the
terminology of \cite{Whisnant,Lin}), i.e. the couplings not present
in the SM Lagrangian ${\cal L}_0$, between the lightest neutral Higgs scalar ($h$)
and the top quark induced by the new physics of  MSSM. We assume that the other
Higgs bosons, as well as all supersymmetric particles, are much heavier
than the lightest neutral Higgs particle, so that the corresponding
fields can be integrated out. A well known fact
is that the masses of the other two neutral Higgs bosons are approximately
equal to that of the charged Higgs boson ($H^+$) under the condition
$m_{_{H^+}}\gg m_{_h}$, and hence one can consider
the lighter Higgs doublet as the SM Higgs field and integrate out
the heavier Higgs doublet.

Our presentation is organized as follows. In Section
II, the notations adopted in this work are introduced.
In Section III we shall describe the method of obtaining the Wilson
coefficients by integrating out the heavy degrees of freedom in
the full theory. The numerical analysis of the constraints on the
parameter space from the present experiments, especially by the
$R_{b}$ data, is given in Section IV. Section V
summarizes our results. Some lengthy formulae, such as the
expressions for the Wilson coefficients and the loop integral
functions, are collected in Appendices.

\section{Preliminaries}

\indent\indent
The most general gauge invariant superpotential, which  retains all
the conservation laws of the SM, is given by
\begin{eqnarray}
{\cal W}=\mu\epsilon_{ij}\hat{H}_{_u}^i\hat{H}_{_d}^j+\epsilon_{ij}
h_{_L}^{JI}\hat{H}_{_d}^i\hat{L}_j^I\hat{R}^J+\epsilon_{ij}
h_{_{D}}^{JI}\hat{H}_{_d}^i\hat{Q}_j^I\hat{D}^J+\epsilon_{ij}
h_{_{U}}^{JI}\hat{H}_{_u}^i\hat{Q}_j^I\hat{U}^J.
 \label{sp}
\end{eqnarray}
Here $\hat{H}_{_u},\;\hat{H}_{_d}$ are the two Higgs superfield doublets,
$\hat{Q}^{I}$ and $\hat{L}^{I}$ are the doublets of quark and lepton superfields, and
$\hat{U}^{I}$,
$\hat{D}^{I}$ and $\hat{R}^{I}$ are the singlet superfields of u-
and d-type quarks and  charged leptons, respectively
(I=1, 2, 3 is generation index, $i,j=1,2$ are SU(2) indices). Yukawa coupling
constants are denoted by $h_{_L}$, $h_{_{U,D}}$. The breaking of supersymmetry
happens through the so-called soft terms, which are in the most general case given by
\begin{eqnarray}
&&{\cal
L}_{soft}=-m_{_{H_u}}^2H_{_u}^{i*}H_{_u}^i-m_{_{H_d}}^2H_{_d}^{i*}H_{_d}^i
-m_{_{L^{IJ}}}^2\tilde{L}_i^{I*}\tilde{L}_i^{J}-m_{_{R^{IJ}}}^2
\tilde{R}^{I*}\tilde{R}^{J}
-m_{_{Q^{IJ}}}^2\tilde{Q}_i^{I*}\tilde{Q}_i^{J}
-m_{_{U^{IJ}}}^2\tilde{U}^{I*}\tilde{U}^{J}\nonumber \\
&&\hspace{1.4cm}-m_{_{D^{IJ}}}^2\tilde{D}^{I*}\tilde{D}^{J}
+(m_1\lambda_B\lambda_B+m_2\lambda_A^i\lambda_A^i
+m_3\lambda_G^a\lambda_G^a+h.c.) +\Big[\epsilon_{ij}m_{_{H_{12}}}^2
H_{_u}^iH_{_d}^j
+\epsilon_{ij}A_{_L}^{JI}H_{_d}^{i}\tilde{L}^{I}_{j}\tilde{R}^{J}
\nonumber \\
&&\hspace{1.4cm}
+\epsilon_{ij}A_{_D}^{JI}H_{_d}^{i}\tilde{Q}^{I}_{j}\tilde{D}^{J}
+\epsilon_{ij}A_{_U}^{JI}H_{_u}^{i}\tilde{Q}^{I}_{j}
\tilde{U}^{J}+h.c.\Big]. \label{soft}
\end{eqnarray}
Here
 $\lambda_G^a\;(a=1,\;2,\;
\cdots\;8),\;\lambda_A^i\;(i=1,\;2,\;3)$ and $\lambda_B$ denote the
$SU(3)$, $SU(2)$ and $U(1)$ gauginos, respectively, and
$A_{_{U,D,L}}^{IJ}$ are coupling constants of the unit of mass.

Let us define scalar doublets $\Phi$ and $\Phi_H$ as follows:
\begin{eqnarray}
&&\left(\begin{array}{l}\Phi\\ \Phi_{_H}\end{array}\right)=\left(
\begin{array}{cc}c_{_\beta}&-s_{_\beta}\\s_{_\beta}&c_{_\beta}
\end{array}\right)\left(\begin{array}{c}\tilde{H}_{_d}\\H_{_u}\end{array}
\right)\;,
\label{smh}
\end{eqnarray}
where $\tilde{H}_{_d}=i\sigma_2H_{_d}^*$ and
$c_{_\beta}=\cos\beta\;,s_{_\beta} =\sin\beta$ with
$\tan\beta={\upsilon_{_u}/\upsilon_{_d}}$, the ratio of the VEVs
of $H_{_u},\;H_{_d}$. With this definition $\Phi$ is identified as
the SM Higgs doublet, consisting of Goldstone bosons and a
physical neutral Higgs field. More explicitly, one can write the
two Higgs doublets as
\begin{eqnarray}
&&\Phi=\left(\begin{array}{c}G^+\\{1\over\sqrt{2}}\Big(\upsilon+H_1^0+iG^0\Big)
\end{array}\right)\;,\;\;\;
\Phi_{_H}=\left(\begin{array}{c}H^+\\{1\over\sqrt{2}}\Big(H_2^0+iA\Big)
\end{array}\right)\; ,\label{ph}
\end{eqnarray}
where $G^0,\;G^+$ denote the Goldstone bosons, $H_1^0$ and $H_2^0$
are the neutral Higgs fields, $H^+$ and $A$ are the physical
charged Higgs and $CP$-odd neutral Higgs bosons, respectively, and
$\upsilon=\sqrt{\upsilon_{_u}^2+\upsilon_{_d}^2} =246\; {\rm
GeV}$.  At the electro-weak scale, the two physical $CP$-even
neutral Higgs fields are obtained through the mixing between the
fields $H_1^0$ and $ \;H_2^0$. The masses of the physical Higgs
bosons are given by
\begin{eqnarray}
&&{\bf m}_{_{even}}^2=\left(\begin{array}{cc}m_{_{\rm Z}}^2\Big(s_{_\beta}^2
-c_{_\beta}^2\Big)^2&2m_{_{\rm Z}}^2s_{_\beta}
c_{_\beta}(c_{_\beta}^2-s_{_\beta}^2)\\2m_{_{\rm Z}}^2s_{_\beta}
c_{_\beta}(c_{_\beta}^2-s_{_\beta}^2)&{m_{_{H_{12}}}^2\over s_{_\beta}
c_{_\beta}}+4m_{_{\rm Z}}^2s_{_\beta}^2c_{_\beta}^2\end{array}\right)
\;\;\;({\rm in\; the\; basis\; (H_1^0,\;H_2^0)^T}),\nonumber\\
&&m_{_A}^2={m_{_{H_{12}}}^2\over s_{_\beta}c_{_\beta}}\;,\nonumber\\
&&m_{_{H^+}}={m_{_{H_{12}}}^2\over s_{_\beta}c_{_\beta}}+m_{_{\rm W}}^2\;.
\label{mass1}
\end{eqnarray}
In the limit $m_{_{H}}^2\gg m_{_{\rm W}}^2$ the two doublets $\Phi$ and $\Phi_H$ decouple,
the former remaining light and the latter being associated with a large
mass $m_{_H}^2={m_{_{H_{12}}}^2/ s_{_\beta}c_{_\beta}}$.

In the following,
we will use the four-component spinor representation for fermions.
From the two-component Wyel spinors
$\psi_{_{Q_I}},\;\psi_{_{U_I}},\;\psi_{_{D_I}},\;\psi_{_{H_u}}$
and $\psi_{_{H_d}}$, we form the following four-component Dirac
fermions:
\begin{eqnarray}
&&q_{_L}^I=\left(\begin{array}{c}\psi_{_{Q_I}}\\0\end{array}\right)\;,\;\;\;
u_{_R}^I=\left(\begin{array}{c}0\\ \bar{\psi}_{_{U_I}}\end{array}\right)\;,
\nonumber\\
&&d_{_R}^I=\left(\begin{array}{c}0\\ \bar{\psi}_{_{D_I}}\end{array}\right)\;,\;\;\;
\psi_{_H}=\left(\begin{array}{c}\psi_{_{H_u}}\\ \overline{\tilde{\psi}}_{_{H_d}}
\end{array}\right)\;,
\label{spinor1}
\end{eqnarray}
with $\tilde{\psi}_{_{H_d}}=\Big(i\sigma^2\Big)\psi_{_{H_d}}$.
Similarly, for the $SU(3)\times SU(2)\times U(1)$ gauginos
$\lambda_G^a, \;\lambda_{A}^i,\;\lambda_B$ can we define the
following four-component Majorana spinors:
\begin{eqnarray}
&&\psi_{_{G}}^a=\left(\begin{array}{c}i\lambda_{G}^a\\ -i\bar{\lambda}_G^a
\end{array}\right)\;,\;\;\;\psi_{_{A}}^i=\left(\begin{array}{c}i\lambda_{A}^i
\\ -i\bar{\lambda}_A^i\end{array}\right)\nonumber\\
&&\psi_{_{B}}=\left(\begin{array}{c}i\lambda_{B}\\ -i\bar{\lambda}_B
\end{array}\right)\;.
\label{spinor2}
\end{eqnarray}

Diagonalizing the soft mass terms is done with  help of the
sfermion mixing matrices  ${\cal Z}_{_{Q,U,D}}$ defined as
\begin{eqnarray}
&&{\cal Z}_{_Q}^\dagger{\bf m}^2_{_{Q}}{\cal Z}_{_Q}=\hat{\bf m}^2_{_{Q}}
\;,\nonumber\\
&&{\cal Z}_{_U}^\dagger{\bf m}^2_{_{U}}{\cal Z}_{_U}=\hat{\bf m}^2_{_{U}}
\;,\nonumber\\
&&{\cal Z}_{_D}^\dagger{\bf m}^2_{_{D}}{\cal Z}_{_D}=\hat{\bf
m}^2_{_{D}} \;, \label{mix1}
\end{eqnarray}
where the matrices $\hat{\bf m}^2_{_{Q,U,D}}$
on the right-handed side are diagonal.

Finally, we will benefit in our calculations from the
following   rearrangement identities of the $SU(2)$ group indices:
\begin{eqnarray}
&&{\bf 1}_{\alpha\alpha^\prime}\otimes{\bf 1}_{\beta\beta^\prime}=
{1\over2}\Big\{{\bf 1}_{\alpha\beta}\otimes{\bf 1}_{\beta^\prime\alpha^\prime}
+\sum\limits_a\sigma^a_{\alpha\beta}\otimes\sigma^a_{\beta^\prime\alpha^\prime}
\Big\}\;,\nonumber\\
&&\sigma^a_{\alpha\alpha^\prime}\otimes\sigma^a_{\beta\beta^\prime}=
{1\over2}\Big\{3{\bf 1}_{\alpha\beta}\otimes{\bf 1}_{\beta^\prime\alpha^\prime}
-\sum\limits_a\sigma^a_{\alpha\beta}\otimes\sigma^a_{\beta^\prime\alpha^\prime}
\Big\}\;,\nonumber\\
&&\sum\limits_{a,b}\epsilon_{abc}\epsilon_{abd}=2\delta_{cd}\;.
\label{resu2}
\end{eqnarray}

\section{The Higgs-top anomalous couplings}

\indent\indent
In this section we shall discuss the anomalous couplings
of the top quark and Higgs bosons. Considering the suppression of
the new physics energy scale, we just keep  operators up to
dimension-six in the effective Lagrangian Eq.  (\ref{elag}). The
top-Higgs anomalous couplings of interest can be classified into
three types: the anomalous couplings involving
a left-handed quark, the right-handed top quark and Higgs boson (${\cal
O}_{_{tq\Phi}}$), the couplings between the Higgs boson and a left-handed
quark (${\cal O}_{_{q\Phi}}$), and the couplings between the Higgs boson and
the right-handed top quark (${\cal O}_{_{t\Phi}}$). After the EW
symmetry breaking, these operators produce not only corrections to
the effective couplings $Wt\bar{b},\; Xt\bar{t},\;Xb\bar{b}$
($X=\gamma,\;Z,\;H$),  but also induce anomalous couplings such as
$\gamma Ht\bar{t},\;ZHt\bar{t}$. All those effects may be detectable
at the Next Linear Collider (NLC) and at the  Tevatron.

In the following subsections we will give the explicit expressions for the
contributions of supersymmetric particles and the heavy Higgs boson
doublet to the effective operators mentioned above by deriving the
relevant Wilson coefficients.  We will give our results in terms of
the following loop integral functions:
\begin{eqnarray}
&&{\cal B}_{_{j,k}}^i(x_{_a},x_{_b})={(4\pi)^2\over i}\int{d^4q\over(2\pi)^4}
{\Big(q^2\Big)^i\over(q^2-x_{_a})^j(q^2-x_{_b})^k}\;,\nonumber\\
&&{\cal C}_{_{jkl}}^i(x_{_a},x_{_b},x_{_c})={(4\pi)^2\over i}
\int{d^4q\over(2\pi)^4}{\Big(q^2\Big)^i\over(q^2-x_{_a})^j(q^2-x_{_b})^k
(q^2-x_{_c})^l}\;,\nonumber\\
&&{\cal D}_{_{jklm}}^i(x_{_a},x_{_b},x_{_c},x_{_d})={(4\pi)^2\over
i}
\int{d^4q\over(2\pi)^4}{\Big(q^2\Big)^i\over(q^2-x_{_a})^j(q^2-x_{_b})^k
(q^2-x_{_c})^l(q^2-x_{_d})^m}\;. \label{loopfun}
\end{eqnarray}
The  explicit expressions  of these are given in Ref. \cite{fengtf}.

\subsection{The anomalous couplings ${\cal O}_{_{tq\Phi}}$ \label{sub1}}

\indent\indent This class of operators includes the $CP$-even
operators
\begin{eqnarray}
&&{\cal O}_{_{tq\Phi1}}=\Big(\Phi^\dagger\Phi\Big)\Big(\bar{q}_{_L}t_{_R}\tilde{\Phi}
+\tilde{\Phi}^\dagger\bar{t}_{_R}q_{_L}\Big)\;,\nonumber\\
&&{\cal O}_{_{tq\Phi2}}=\bar{q}_{_L}\Big(D_\mu t_{_R}\Big)D^\mu\tilde{\Phi}
+\Big(D_\mu\tilde{\Phi}\Big)^\dagger\overline{\Big(D^\mu t_{_R}\Big)}q_{_L}
\;,\nonumber\\
&&{\cal O}_{_{tq\Phi3}}=\overline{\Big(D_\mu q_{_L}\Big)}\Big(D^\mu t_{_R}\Big)\tilde{\Phi}
+\tilde{\Phi}^\dagger\overline{\Big(D_\mu t_{_R}\Big)}\Big(D^\mu q_{_L}\Big)
\;,\nonumber\\
&&{\cal O}_{_{tq\Phi4}}=\overline{\Big(D_\mu q_{_L}\Big)}t_{_R}\Big(D^\mu \tilde{\Phi}\Big)
+\Big(D_\mu \tilde{\Phi}\Big)^\dagger\bar{t}_{_R}\Big(D^\mu q_{_L}\Big)
\;,\nonumber\\
&&{\cal O}_{_{tq\Phi5}}=i\overline{\Big(D_\mu q_{_L}\Big)}\sigma^{\mu\nu}
t_{_R}\Big(D_\nu \tilde{\Phi}\Big)
+i\Big(D_\nu \tilde{\Phi}\Big)^\dagger\bar{t}_{_R}
\sigma^{\mu\nu}\Big(D_\mu q_{_L}\Big)
\;,\nonumber\\
\label{op1e}
\end{eqnarray}
where the covariant derivative $D_{_\mu}$ is given by
$D_{_\mu}=\partial_{_\mu}-{i\over2}g_{_3}T^{^A}G^{^A}_\mu-{i\over2}g_{_2}
\sigma^aW^a_\mu-{i\over2}g_{_1}Y_{_B}B_\mu$. The
$CP$-odd counterparts of these operators are
\begin{eqnarray}
&&{\cal O}_{_{tq\Phi6}}=\Big(\Phi^\dagger\Phi\Big)\Big(\bar{q}_{_L}t_{_R}\tilde{\Phi}
-\tilde{\Phi}^\dagger\bar{t}_{_R}q_{_L}\Big)\;,\nonumber\\
&&{\cal O}_{_{tq\Phi7}}=\bar{q}_{_L}\Big(D_\mu t_{_R}\Big)D^\mu\tilde{\Phi}
-\Big(D_\mu\tilde{\Phi}\Big)^\dagger\overline{\Big(D^\mu t_{_R}\Big)}q_{_L}
\;,\nonumber\\
&&{\cal O}_{_{tq\Phi8}}=\overline{\Big(D_\mu q_{_L}\Big)}\Big(D^\mu t_{_R}\Big)\tilde{\Phi}
-\tilde{\Phi}^\dagger\overline{\Big(D_\mu t_{_R}\Big)}\Big(D^\mu q_{_L}\Big)
\;,\nonumber\\
&&{\cal O}_{_{tq\Phi9}}=\overline{\Big(D_\mu q_{_L}\Big)}t_{_R}\Big(D^\mu \tilde{\Phi}\Big)
-\Big(D_\mu \tilde{\Phi}\Big)^\dagger\bar{t}_{_R}\Big(D^\mu q_{_L}\Big)
\;,\nonumber\\
&&{\cal O}_{_{tq\Phi10}}=i\overline{\Big(D_\mu
q_{_L}\Big)}\sigma^{\mu\nu} t_{_R}\Big(D_\nu \tilde{\Phi}\Big)
-i\Big(D_\nu \tilde{\Phi}\Big)^\dagger\bar{t}_{_R}
\sigma^{\mu\nu}\Big(D_\mu q_{_L}\Big) \;. \label{op1o}
\end{eqnarray}

\begin{figure}
\setlength{\unitlength}{1mm}
\begin{center}
\begin{picture}(230,200)(50,90)
\put(50,80){\includegraphics{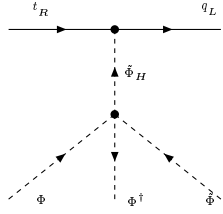}}
\end{picture}
\caption[]{The Feynman diagrams inducing nontrivial contributions
to the Wilson coefficients of the operators ${\cal
O}_{_{tq\Phi1}}$ and ${\cal O}_{_{tq\Phi6}}$} \label{fig1}
\end{center}
\end{figure}

For the $CP$-even operator ${\cal O}_{_{tq\Phi1}}$ and for the
corresponding $CP$-odd operator ${\cal O}_{_{tq\Phi6}}$,
nonzero contributions to the Wilson coefficients originate
from the Feynman diagrams shown in Fig.  \ref{fig1}, and they are given by
\begin{eqnarray}
&&C_{_{tq\Phi1}}={1\over2}\Big(g_1^2+g_2^2\Big){\bf Re}(
{\bf h}_{_U}^{33}){s_{_\beta}c_{_\beta}^2(c_{_\beta}^2-s_{_\beta}^2)
\over x_{_H}}\;,\nonumber\\
&&C_{_{tq\Phi6}}=i{1\over2}\Big(g_1^2+g_2^2\Big){\bf Im}(
{\bf h}_{_U}^{33}){s_{_\beta}c_{_\beta}^2(c_{_\beta}^2-s_{_\beta}^2)
\over x_{_H}}\;,
\label{wcoe1}
\end{eqnarray}
where $x_{_H}={m_{_H}^2/ \mu_{_{NP}}^2}$. In the full theory,
the Feynman diagrams that induce the anomalous couplings ${\cal
O}_{_{tq\Phi1}},\;{\cal O}_{_{tq\Phi6}}$ should also include
diagrams involving virtual SM fields. However, these diagrams have no contribution to the
Wilson coefficients after matching  the effective Lagrangian
Eq. (\ref{elag}) with the Lagrangian of the full theory (MSSM) (see below for more
details).

\begin{figure}
\setlength{\unitlength}{1mm}
\begin{center}
\begin{picture}(100,130)(85,110)
\put(50,80){\includegraphics{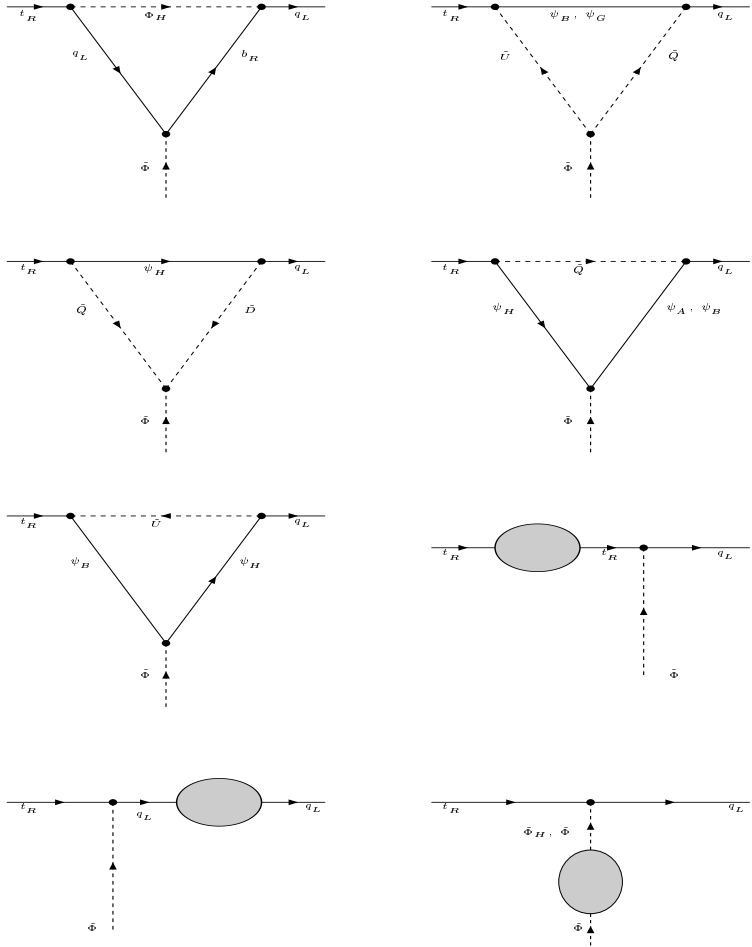}}
\end{picture}
\caption[]{The Feynman diagrams inducing nontrivial contribution
to the Wilson coefficients of the operators ${\cal O}_{_{tq\Phi
i}}\;(i= 2,\;\cdots,\;5,\;7\;\cdots,\;10)$ in the full theory.}
\label{fig2}
\end{center}
\end{figure}

For the $CP$-even anomalous operators ${\cal O}_{_{tq\Phi
i}}\;(i=2,\;3,\;4,\;5)$ and the $CP$-odd anomalous operators
${\cal O}_{_{tq\Phi i}} \;(i=7,\;8,\;9,\;10)$, the derivation of
the Wilson coefficients leads to relatively tedious calculations.
In  Fig.  \ref{fig2} we show the Feynman diagrams, which induce
nontrival contributions to the Wilson coefficients after matching
the amplitude of the effective theory with that of the MSSM. In
these diagrams, the black blobs represent the self-energy diagrams
of $\bar{q}_{_L}q_{_L}$, $\bar{t}_{_R}t_{_R}$, and
$\Phi^\dagger\Phi\;(\Phi_H)$ (Fig. \ref{fig4}).

The matching procedure, to which we refer above, is extensively applied in the derivation of
the effective Lagrangian in the hadron physics, especially in
the application of the effective Lagrangian to the rare $B$
decay  \cite{Buras}. The main idea of this procedure is the following.
We derive the amplitude corresponding to the relevant Feynman diagrams both in the
full theory and in the effective theory. In both derivations we
only keep the momenta $p_i$ of external
particles to the second order. Through a comparison of the amplitudes of
the full theory and the effective theory we then obtain the
Wilson coefficients of interest.

For a demonstration, let us consider the  first diagram of Fig.
\ref{fig2}.  In the full theory we can write the amplitude
corresponding to this diagram as
\begin{eqnarray}
&&A_{_{2(1)}}^{FT}(p,q)=-{i\over(4\pi)^2}s_{_\beta}c_{_\beta}^2\Big({\bf h}_{_D}^\dagger
{\bf h}_{_D}{\bf h}_{_U}^\dagger\Big)_{33}\bigg\{\Big[\Delta+1+\ln{\mu_{_{NP}}^2\over
m_{_H}^2}\Big]\Big(\bar{q}_{_L}\tilde{\Phi}\Big)t_{_R}
-{1\over2m_{_H}^2}\Big(1+\ln{m_{_q}^2\over
m_{_H}^2}\Big)\Big(\bar{q}_{_L}\tilde{\Phi}\Big)q^2t_{_R}
\nonumber\\&&\hspace{2.2cm}
+{1\over2m_{_H}^2}\Big(\bar{q}_{_L}\tilde{\Phi}\Big)(p+q)^2t_{_R}
-{1\over2m_{_H}^2}\Big(\bar{q}_{_L}\tilde{\Phi}\Big)q\cdot(p+q)t_{_R}
+{1\over4m_{_H}^2}\Big(\bar{q}_{_L}\tilde{\Phi}\Big)
[/\!\!\!p,/\!\!\!q]t_{_R}\bigg\}\;.
\label{match1}
\end{eqnarray}
Here $\Delta={1\over\epsilon}-\gamma_{_E}+\ln4\pi$ denotes the
ultraviolet divergence ($D=4-2\epsilon$ is the time-space
dimension in the dimensional regularization scheme),
$\mu_{_{NP}}$ is the scale of new physics, and  $p$ and $q$ denote the
four-momenta of the external particles $t_{_R}$ and $\Phi$,
respectively.
 In the full theory the light
fields and the heavy fields co-exist in the Lagrangian. When the
heavy degrees of freedom are integrated out and
the light fields are treated as massless, infrared divergences are encountered.
They are regulated by the parameter $m_{_q}$.

\begin{figure}
\setlength{\unitlength}{1mm}
\begin{center}
\begin{picture}(80,30)(90,230)
\put(50,80){\includegraphics{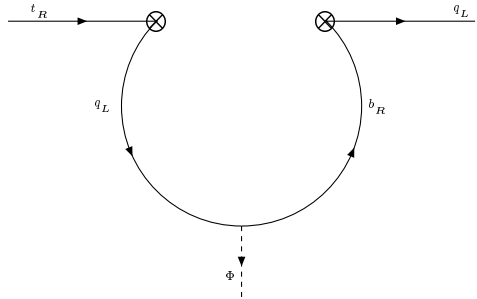}}
\end{picture}
\caption[]{The Feynman diagram corresponds to the first diagram
of Fig. \ref{fig2} in the effective theory.}
\label{fig3}
\end{center}
\end{figure}

\begin{figure}
\setlength{\unitlength}{1mm}
\begin{center}
\begin{picture}(100,50)(70,240)
\put(50,80){\includegraphics{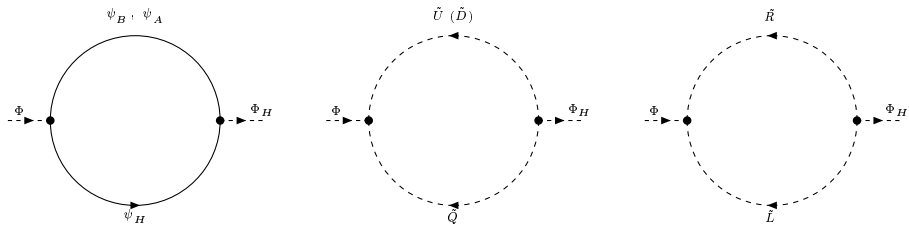}}
\end{picture}
\caption[]{The Higgs self-energy diagrams which induce nonzero
contributions to the Wilson coefficients of operators
Eq. (\ref{op1e}) and Eq. (\ref{op1o}).} \label{fig4}
\end{center}
\end{figure}

The amplitude of the corresponding Feynman diagram
in the effective theory, presented in Fig. \ref{fig3}, is given by
\begin{eqnarray}
&&A_{_{2(1)}}^{ET}(p,q)=-{i\over2(4\pi)^2m_{_H}^2}s_{_\beta}c_{_\beta}^2
\Big({\bf h}_{_D}^\dagger {\bf h}_{_D}{\bf h}_{_U}^\dagger\Big)_{33}
\Big(\Delta-{1\over2}+\ln{\mu_{_{NP}}^2\over m_{_q}^2}
\Big)\Big(\bar{q}_{_L}\tilde{\Phi}\Big)q^2t_{_R}\;.
\label{match2}
\end{eqnarray}

In the operators of the effective theory, only the light fields
exist, and the Wilson coefficients do not depend on their masses.
As in the full theory, an infrared divergence emerges  here, and
it is also regularized by the parameter $m_{_q}$. As expected, the
infrared divergence appearing in the effective theory is the same
as that appearing in the full-theory. By matching the amplitudes
Eq. (\ref{match2}) and Eq. (\ref{match1}), one gets rid of the
infrared divergence. After this matching step, we can present the
amplitude in its final form:

\begin{eqnarray}
&&A_{_{2(1)}}(p,q)=-{i\over(4\pi)^2}s_{_\beta}c_{_\beta}^2\Big({\bf h}_{_D}^\dagger
{\bf h}_{_D}{\bf h}_{_U}^\dagger\Big)_{33}\bigg\{\Big[\Delta+1+\ln{\mu_{_{NP}}^2\over
m_{_H}^2}\Big]\Big(\bar{q}_{_L}\tilde{\Phi}\Big)t_{_R}
+{1\over2m_{_H}^2}\Big(1+\ln{\mu_{_{NP}}^2\over
m_{_H}^2}\Big)\Big(\bar{q}_{_L}\tilde{\Phi}\Big)p\cdot qt_{_R}
\nonumber\\&&\hspace{2.0cm}
-{1\over2m_{_H}^2}\Big(1+\ln{\mu_{_{NP}}^2\over
m_{_H}^2}\Big)\Big(\bar{q}_{_L}\tilde{\Phi}\Big)q\cdot(p+q)t_{_R}
+{1\over2m_{_H}^2}\Big(\bar{q}_{_L}\tilde{\Phi}\Big)p\cdot(p+q)t_{_R}
\nonumber\\&&\hspace{2.0cm}
-{1\over4m_{_H}^2}\Big(\bar{q}_{_L}\tilde{\Phi}\Big)
[/\!\!\!q,/\!\!\!p]t_{_R}\bigg\}\;.
\label{match3}
\end{eqnarray}

The first term in the parenthesis of Eq. (\ref{match3}) contributes to
the renormalization of the Yukawa couplings ${\bf h}_{_U}$, and it is
irrelevant to our present discussion, taking into account the approximation
level we work on. For those diagrams where the inner lines are
supersymmetry particles, the Wilson
coefficients of the anomalous couplings can be directly read from the amplitudes,
because we integrate out all the supersymmetry fields in the
effective theory.

Now we will turn to show how to obtain the contributions of the
self-energy diagrams to the anomalous couplings. As mentioned above,
there are three possible self-energy diagrams that contribute to the
coefficients indirectly, namely the self-energy corrections to
the $\bar{q}_{_L}q_{_L}$, $\bar{t}_{_R}t_{_R}$ , and Higgs doublet
currents. For a fermion, the renormalized fields are
defined by
\begin{eqnarray}
&&f_{_{L,i}}^0=Z_{_{L,ij}}^{1\over2}f_{_{L,j}}\;,\nonumber\\
&&f_{_{R,i}}^0=Z_{_{R,ij}}^{1\over2}f_{_{R,j}}\;,
\label{renf}
\end{eqnarray}
where $i,j$ are generation indices, $f_{_{L,i}}^0,\;f_{_{R,i}}^0$
are the left- and right-handed bare fields, respectively,
$f_{_{L,i}},\;f_{_{R,i}}$ are the corresponding renormalized
fields, and $Z_{_{L,R}}$ are the wave function renormalization
constants. Ignoring the fermion masses, we can write down the
counter terms for the fermions in Eq. (\ref{renf}) as follows:
\begin{eqnarray}
&&\Sigma_{_{ij}}^{L,c}(p)=\Big(Z_{_{L,iI}}^{\dagger{1\over2}}
Z_{_{L,Ij}}^{{1\over2}}-\delta_{_{ij}}\Big)/\!\!\!p={1\over2}\Big(\delta Z_{_{L,ij}}^\dagger
+\delta Z_{_{L,ij}}\Big)/\!\!\!p\;,\nonumber\\
&&\Sigma_{_{ij}}^{R,c}(p)=\Big(Z_{_{R,iI}}^{\dagger{1\over2}}
Z_{_{R,Ij}}^{{1\over2}}-\delta_{_{ij}}\Big)/\!\!\!p={1\over2}\Big(\delta Z_{_{R,ij}}^\dagger
+\delta Z_{_{R,ij}}\Big)/\!\!\!p\;,
\label{couf}
\end{eqnarray}
where $p$ denotes the external momentum of the fermion. In the
full theory, we  express the bare self-energy of the fermions as
\begin{eqnarray}
&&\Sigma_{_{ij}}^{L,0}=\Big[\delta_{_{ij}}+A_{_{ij}}^L
+B_{_{ij}}^Lp^2\Big]/\!\!\!p\;,\nonumber\\
&&\Sigma_{_{ij}}^{R,0}=\Big[\delta_{_{ij}}+A_{_{ij}}^R
+B_{_{ij}}^Rp^2\Big]/\!\!\!p
\label{fbself}
\end{eqnarray}
where the first term $\delta_{_{ij}}$ represents the Born
approximation part and  $A^{L,R},\;B^{L,R}$  originate from
raditive corrections. From  Eq. (\ref{couf}) and Eq.
(\ref{fbself}) one finds the following form for the renormalized
self-energies:
\begin{eqnarray}
&&\hat{\Sigma}_{_{ij}}^L=\Big[\delta_{_{ij}}+{1\over2}\Big(\delta Z_{_{L,ij}}^\dagger
+\delta Z_{_{L,ij}}\Big)+A_{_{ij}}^L+B_{_{ij}}^Lp^2\Big]/\!\!\!p\;,\nonumber\\
&&\hat{\Sigma}_{_{ij}}^R=\Big[\delta_{_{ij}}+{1\over2}\Big(\delta Z_{_{R,ij}}^\dagger
+\delta Z_{_{R,ij}}\Big)+A_{_{ij}}^R+B_{_{ij}}^Rp^2\Big]/\!\!\!p\;.
\label{frself}
\end{eqnarray}

The explicit expressions of the renormalization constant $\delta
Z_{_{L,R}}$ depend upon the renormalization scheme, {\it i.e.},
the renormalization conditions. Instead of the often-used
renormalization schemes, i.e. the minimal subtraction scheme
($MS$) or the modified minimal subtraction scheme
($\overline{MS}$), we adopt here the physical renormalization
conditions
\begin{eqnarray}
&&\hat{\Sigma}_{_{ij}}^Lf_{_{L,j}}|_{_{/\!\!\!p=0}}=0\;,\nonumber\\
&&\hat{\Sigma}_{_{ij}}^Rf_{_{R,j}}|_{_{/\!\!\!p=0}}=0\;,\nonumber\\
&&{1\over /\!\!\!p}\hat{\Sigma}_{_{ij}}^Lf_{_{L,j}}|_{_{/\!\!\!p=0}}
=f_{_{L,i}}\;,\nonumber\\
&&{1\over /\!\!\!p}\hat{\Sigma}_{_{ij}}^Rf_{_{R,j}}|_{_{/\!\!\!p=0}}
=f_{_{R,i}}\;.\nonumber\\
\label{condition1}
\end{eqnarray}
The first two conditions mean that the renormalized fields
satisfy the equations of motion of free particles (for
massless fermions this is a trivial constraint), and the last two
conditions set the residue of the propagators at the pole
equal to unity. In fact, this scheme is just the on-shell
renormalization scheme often used when calculating radiative corrections to electroweak
processes  \cite{onshell}. Of course, for high
energy processes we can ignore the fermion mass in our
approximation. Using the condition Eq. (\ref{condition1}), we
achieve the renormalized fermion self-energies:
\begin{eqnarray}
&&\hat{\Sigma}_{_{ij}}^L=B_{_{ij}}^Lp^2/\!\!\!p\;,\nonumber\\
&&\hat{\Sigma}_{_{ij}}^R=B_{_{ij}}^Rp^2/\!\!\!p\;.
\label{fself}
\end{eqnarray}
We can attribute those terms to the contributions of the high
dimension operators $\bar{q}_{_L}\Big(i/\!\!\!\!
D\Big)^3q_{_L},\;\bar{t}_{_R} \Big(i/\!\!\!\! D\Big)^3t_{_R}$.
After the matching of the full and effective theories, there is no
contribution to the operators of our interesting given in Eq. (\ref{op1e}) and
Eq. (\ref{op1o}) from the fermion self-energy diagrams.

For the Higgs sector, the bare self-energies are given as
\begin{eqnarray}
&&\Sigma_{_{\Phi\Phi}}^0(p^2)=D_{_{\Phi\Phi}}+\Big(1+E_{_{\Phi\Phi}}\Big)p^2
+F_{_{\Phi\Phi}}p^4\;,\nonumber\\
&&\Sigma_{_{\Phi\Phi_H}}^0(p^2)=D_{_{\Phi\Phi_H}}+E_{_{\Phi\Phi_H}}p^2
+F_{_{\Phi\Phi_H}}p^4\;,
\label{bbself}
\end{eqnarray}
where $p$  denotes the momentum of the external particle. In Eq.
(\ref{bbself}),  $D,\;E,\;F$ are standard integral functions that
appear in radiative corrections. For the renormalization of the
Higgs boson wave function and mass, we request the renormalized
boson self-energy to satisfy the conditions
\begin{eqnarray}
&&\hat{\Sigma}_{_{\Phi\Phi}}(p^2)|_{_{p^2=0}}=0\;,\nonumber\\
&&{1\over p^2}\hat{\Sigma}_{_{\Phi\Phi}}(p^2)|_{_{p^2=0}}=0\;,\nonumber\\
&&\hat{\Sigma}_{_{\Phi\Phi_H}}(p^2)|_{_{p^2=0}}=0\;,\nonumber\\
&&\hat{\Sigma}_{_{\Phi\Phi_H}}(p^2)|_{_{p^2=m_{_H}^2}}=0\;.
\label{condition2}
\end{eqnarray}
It is easy to find the renormalized Higgs field self-energies
which meet the conditions of Eq. (\ref{condition2}):
\begin{eqnarray}
&&\hat{\Sigma}_{_{\Phi\Phi}}(p^2)=F_{_{\Phi\Phi}}p^4\;,\nonumber\\
&&\hat{\Sigma}_{_{\Phi\Phi_H}}(p^2)=F_{_{\Phi\Phi_H}}p^2
(p^2-m_{_H}^2)\;.
\label{brself}
\end{eqnarray}
The function $F_{_{\Phi\Phi}}$ is attributed to the contribution
of the high dimensional operator $\Phi^\dagger\Big(D_\mu
D^\mu\Big)^2\Phi$. After the matching procedure, this piece will
not contribute to the operators in (\ref{op1e}) and (\ref{op1o})
which we are interested in. In fact, after the matching the only
nonvanishing contributions from the self-energy diagrams to these
operators  originate from the integral function
$F_{_{\Phi\Phi_H}}$, because we integrate the heavy Higgs doublet
out in the effective theory.

After these  preparations, we can now derive the Wilson coefficients of
the operators ${\cal O}_{_{tq\Phi i}}\;(i=2,\;3,\;4,\;5)$ and $
{\cal O}_{_{tq\Phi i}}\;(i=7,\;8,\;9,\;10)$. For clarity, we present their lengthy
expressions in Appendix \ref{aop1}.

\subsection{The anomalous couplings ${\cal O}_{_{t\Phi}}$}

\begin{figure}
\setlength{\unitlength}{1mm}
\begin{center}
\begin{picture}(230,200)(50,90)
\put(50,80){\includegraphics{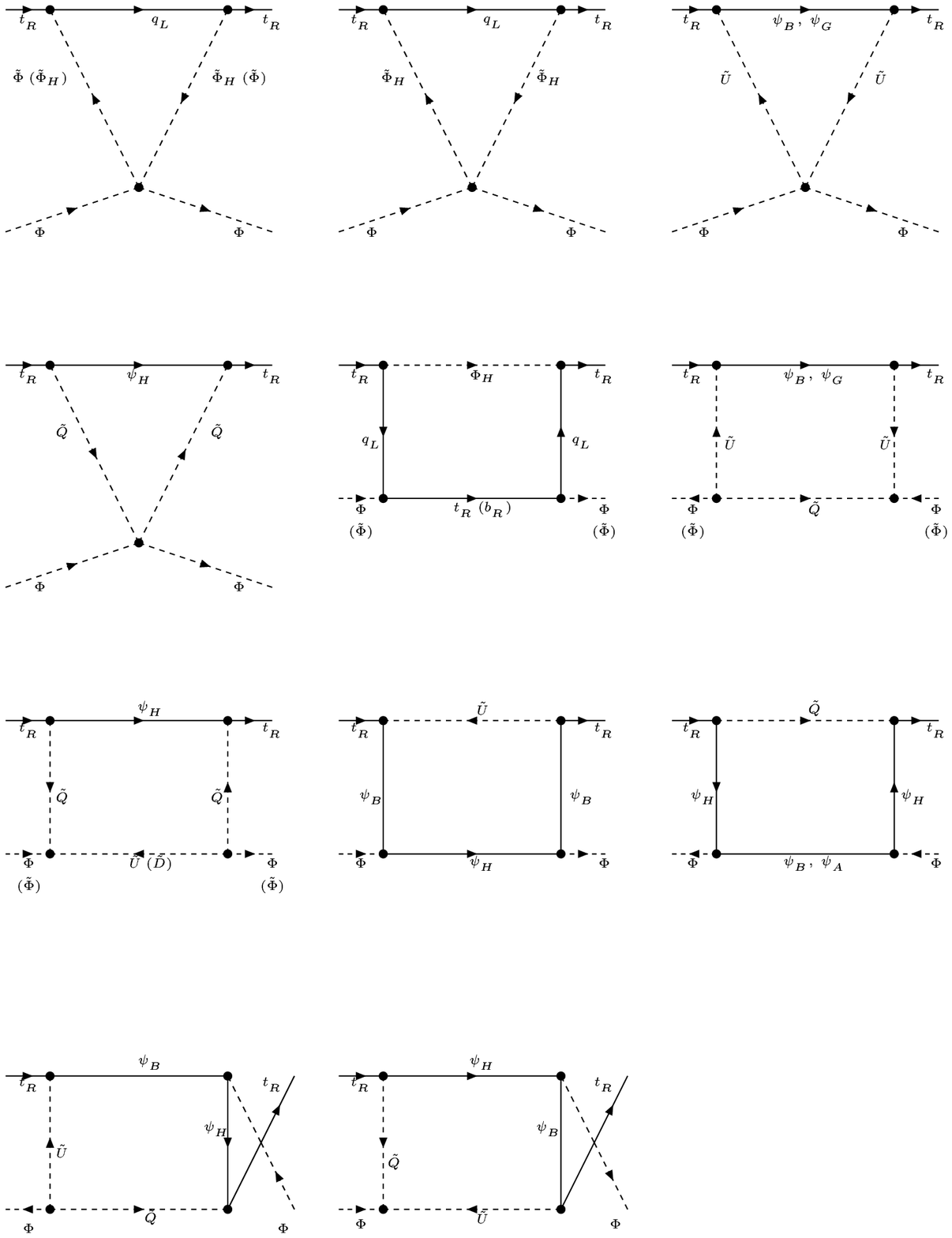}}
\end{picture}
\caption[]{The Feynman diagrams which induce nontrivial
contributions to the Wilson coefficients of the operators ${\cal
O}_{_{t\Phi i}}\;(i= 1,\;2,\;3)$.} \label{fig5}
\end{center}
\end{figure}

\indent\indent This class of anomalous couplings includes the effective
operators
\begin{eqnarray}
&&{\cal O}_{_{t\Phi1}}=i\Big(\Phi^\dagger D_{_\mu}\Phi-(D_{_\mu}\Phi)^\dagger
\Phi\Big)\bar{t}_{_R}\gamma_\mu t_{_R}\;,\nonumber\\
&&{\cal O}_{_{t\Phi2}}=i\Big(\Phi^\dagger\Phi\Big)\Big(\bar{t}_{_R}\gamma^\mu
(D_{_\mu}t_{_R})-(\overline{D_{_\mu}t_{_R}})\gamma^\mu t_{_R}\Big)\;,\nonumber\\
&&{\cal O}_{_{t\Phi3}}=i\Big(\Phi^\dagger D_{_\mu}\Phi+(D_{_\mu}\Phi)^\dagger
\Phi\Big)\bar{t}_{_R}\gamma_\mu t_{_R}\;,
\label{op2}
\end{eqnarray}
where the operators ${\cal O}_{_{t\Phi1}},\;{\cal O}_{_{t\Phi2}}$
have even $CP$-parity and ${\cal O}_{_{t\Phi3}}$ has odd
$CP$-parity. In  Fig. \ref{fig5}, we present
those Feynman diagrams, which induce  nontrivial contributions to
the Wilson coefficients when matching the amplitude obtained in the
effective theory  with that in the full theory (MSSM).
The ensuing Wilson coefficients are collected in
Appendix \ref{aop2}.

\begin{figure}
\setlength{\unitlength}{1mm}
\begin{center}
\begin{picture}(230,200)(50,90)
\put(50,80){\includegraphics{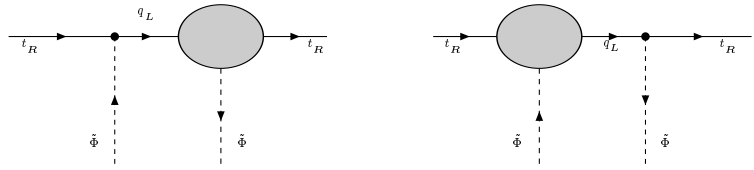}}
\end{picture}
\caption[]{The one-particle-irreducible (1PI) Feynman diagrams
which are related to the Wilson coefficients of the operators
${\cal O}_{_{t\Phi i}}\;(i= 1,\;2,\;3)$ in the full theory, the
gray bulbs represent the diagrams of Fig. \ref{fig2}.}
\label{fig6}
\end{center}
\end{figure}

\begin{figure}
\setlength{\unitlength}{1mm}
\begin{center}
\begin{picture}(230,200)(50,90)
\put(50,80){\includegraphics{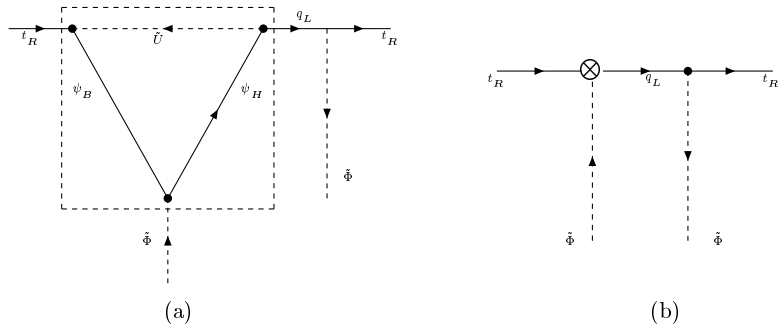}}
\end{picture}
\caption[]{The 1PI Feynman diagrams related to the Wilson
coefficients of the operators ${\cal O}_{_{t\Phi i}}\;(i=
1,\;2,\;3)$ in  (a) the full theory, (b) the effective theory.}
\label{fig7}
\end{center}
\end{figure}

In the full theory, we also include the 1PI diagrams
depicted in   Fig. \ref{fig6}, where the gray blobs represent the
corresponding diagrams of Fig. \ref{fig2}. However, the
contributions from these diagrams disappear as a result of the matching of the
effective theory and  full theory amplitudes.
In order to demonstrate this, let us consider an example.
From the subsection \ref{sub1} , we find that the contributions of
the subdiagram (framed by the dashed lines) in Fig. \ref{fig7}(a)
induce the following term to the effective Lagrangian:
\begin{eqnarray}
&&{\cal L}_{_{eff}}^{\prime}={1\over2\mu_{_{NP}}^2}\sum\limits_{\alpha=2}^5
\Big(C_{_{tq\Phi\alpha}}^{\prime}+C_{_{tq\Phi(5+\alpha)}}^{\prime}\Big)
\Big({\cal O}_{_{tq\Phi\alpha}}+{\cal O}_{_{tq\Phi(5+\alpha)}}\Big),
\label{example0}
\end{eqnarray}
where
\begin{eqnarray}
&&C_{_{tq\Phi2}}^{\prime}=-{1\over48\pi^2}g_1^2\sum\limits_I
\Big[\Lambda_{_{U,I}}^{^T}\Big({\cal C}_{_{121}}^1(x_{_\mu},x_{_1},x_{_{U_I}})
-2s_{_\beta}x_{_\mu}{\cal C}_{_{131}}^1(x_{_\mu},x_{_1},x_{_{U_I}})\Big)
\nonumber\\
&&\hspace{1.5cm}
+2c_{_\beta}\Lambda_{_{U,I}}^{^{R,1}}
{\cal C}_{_{131}}^0(x_{_\mu},x_{_1},x_{_{U_I}})\Big]
\;,\nonumber\\
&&C_{_{tq\Phi3}}^{\prime}={1\over24\pi^2}g_1^2\sum\limits_Ix_{_{U_I}}
\Big[s_{_\beta}\Lambda_{_{U,I}}^{^T}{\cal C}_{_{113}}^1(x_{_\mu},x_{_1},x_{_{U_I}})
-c_{_\beta}\Lambda_{_{U,I}}^{^{R,1}}
{\cal C}_{_{113}}^0(x_{_\mu},x_{_1},x_{_{U_I}})\Big]
\;,\nonumber\\
&&C_{_{tq\Phi4}}^{\prime}=-{1\over48\pi^2}g_1^2\sum\limits_I\Big[
\Lambda_{_{U,I}}^{^T}\Big(Q_9(x_{_1},x_{_\mu},x_{_{U_I}})
-2s_{_\beta}Q_4(x_{_\mu},x_{_1},x_{_{U_I}})\Big)
+2c_{_\beta}\Lambda_{_{U,I}}^{^{R,1}}
Q_3(x_{_\mu},x_{_1},x_{_{Q_I}})\Big]
\;,\nonumber\\
&&C_{_{tq\Phi5}}^{\prime}=-{1\over48\pi^2}g_1^2\sum\limits_I
\Lambda_{_{U,I}}^{^T}{\cal C}_{_{112}}^1(x_{_\mu},x_{_1},x_{_{U_I}})
\;,\nonumber\\
&&C_{_{tq\Phi7}}^{\prime}=-{i\over48\pi^2}g_1^2\sum\limits_I
\Big[\Lambda_{_{U,I}}^{^C}\Big({\cal C}_{_{121}}^1(x_{_\mu},x_{_1},x_{_{U_I}})
-2s_{_\beta}x_{_\mu}{\cal C}_{_{131}}^1(x_{_\mu},x_{_1},x_{_{U_I}})\Big)
\nonumber\\
&&\hspace{1.5cm}
+2c_{_\beta}\Lambda_{_{U,I}}^{^{A,1}}
{\cal C}_{_{131}}^0(x_{_\mu},x_{_1},x_{_{U_I}})\Big]
\;,\nonumber\\
&&C_{_{tq\Phi8}}^{\prime}={i\over24\pi^2}g_1^2\sum\limits_Ix_{_{U_I}}
\Big[s_{_\beta}\Lambda_{_{U,I}}^{^C}{\cal C}_{_{113}}^1(x_{_\mu},x_{_1},x_{_{U_I}})
-c_{_\beta}\Lambda_{_{U,I}}^{^{A,1}}
{\cal C}_{_{113}}^0(x_{_\mu},x_{_1},x_{_{U_I}})\Big]
\;,\nonumber\\
&&C_{_{tq\Phi9}}^{\prime}={i\over48\pi^2}g_1^2\sum\limits_I\Big[
\Lambda_{_{U,I}}^{^C}\Big(Q_9(x_{_1},x_{_\mu},x_{_{U_I}})
-2s_{_\beta}Q_4(x_{_\mu},x_{_1},x_{_{U_I}})\Big)
+2c_{_\beta}\Lambda_{_{U,I}}^{^{A,1}}
Q_3(x_{_\mu},x_{_1},x_{_{Q_I}})\Big]
\;,\nonumber\\
&&C_{_{tq\Phi10}}^{\prime}=-{i\over48\pi^2}g_1^2\sum\limits_I
\Lambda_{_{U,I}}^{^C}{\cal C}_{_{112}}^1(x_{_\mu},x_{_1},x_{_{U_I}})\;,
\label{example1}
\end{eqnarray}
and
$x_{_\mu}={|\mu|^2/\mu_{_{NP}}^2},\;x_{_{U_I}}={m_{_{U_I}}^2/\mu_{_{NP}}^2}$
and $x_{_i}={|m_i|^2/\mu_{_{NP}}^2}\;(i=1,\;2,\;3)$. The
definition of the coupling constants $\Lambda_{_{U,I}}$ and
functions $Q_i(x,y,z)$ can be found in Appendix \ref{app4}.

In the effective theory, the amplitude of  Fig. \ref{fig7}(b) is
written as
\begin{eqnarray}
&&A^{ET}(p_1,q_1,q_2)={i\over2}s_{_{\beta}}{\bf h}_{_U}^{3K}\Big(\Phi^\dagger\Phi\Big)
\bigg\{\Big(C_{_{tq\Phi2}}^{\prime}+C_{_{tq\Phi7}}^{\prime}\Big)
\bar{t}_{_R}{1\over/\!\!\!p_1+/\!\!\!q_1}q_1\cdot p_1t_{_R}
\nonumber\\&&\hspace{3.0cm}
+\Big(C_{_{tq\Phi3}}^{\prime}+C_{_{tq\Phi8}}^{\prime}\Big)
\bar{t}_{_R}{1\over/\!\!\!p_1+/\!\!\!q_1}p_1\cdot(q_1+p_1)t_{_R}
\nonumber\\&&\hspace{3.0cm}
+\Big(C_{_{tq\Phi4}}^{\prime}+C_{_{tq\Phi9}}^{\prime}\Big)
\bar{t}_{_R}{1\over/\!\!\!p_1+/\!\!\!q_1}q_1\cdot(q_1+p_1)t_{_R}
\nonumber\\&&\hspace{3.0cm}
-{1\over2}\Big(C_{_{tq\Phi5}}^{\prime}+C_{_{tq\Phi10}}^{\prime}\Big)
\bar{t}_{_R}{1\over/\!\!\!p_1+/\!\!\!q_1}[/\!\!\!q_1,/\!\!\!p_1]t_{_R}
\bigg\}
\label{example2}
\end{eqnarray}
where $p_1,\;q_1,\;q_2$ denote the four-momenta of the initial
right-handed top quark and the Higgs bosons, respectively. In the
full theory, the corresponding amplitude obtains the form
\begin{eqnarray}
&&A^{FT}(p_1,q_1,q_2)=-{i\over96\pi^2\mu_{_{NP}}^2}s_{_\beta}g_1^2\Big({\bf
h}_{_U} {\bf h}_{_U}^\dagger{\cal Z}_{_U}^\dagger\Big)^{3I}{\cal
Z}_{_U}^{I3}
\Big(\Phi^\dagger\Phi\Big)\bigg\{4\Big[c_{_\beta}\mu_{_{NP}}^2
\sqrt{x_{_\mu}x_{_1}}e^{i(\varphi_1+\varphi_\mu)} {\cal
C}_{_{111}}^0(x_{_\mu},x_{_1},x_{_{U_I}})
\nonumber\\&&\hspace{3.0cm} -s_{_\beta}{\cal
C}_{_{111}}^1(x_{_\mu},x_{_1},x_{_{U_I}})
\Big]\bar{t}_{_R}{1\over/\!\!\!p_1+/\!\!\!q_1}t_{_R}
\nonumber\\&&\hspace{3.0cm} +\Big[2{\cal
C}_{_{121}}^1(x_{_\mu},x_{_1},x_{_{U_I}}) +2{\cal
C}_{_{112}}^1(x_{_\mu},x_{_1},x_{_{U_I}})
\nonumber\\&&\hspace{3.0cm}
+4c_{_\beta}\sqrt{x_{_\mu}x_{_1}}e^{i(\varphi_1+\varphi_\mu)}
Q_3(x_{_\mu},x_{_1},x_{_{U_I}}) \nonumber\\&&\hspace{3.0cm}
-4s_{_\beta}Q_4(x_{_\mu},x_{_1},x_{_{U_I}})
\Big]\bar{t}_{_R}{1\over/\!\!\!p_1+/\!\!\!q_1}q_1\cdot(q_1+p_1)t_{_R}
\nonumber\\&&\hspace{3.0cm} -\Big[2{\cal
C}_{_{121}}^1(x_{_\mu},x_{_1},x_{_{U_I}})
+4c_{_\beta}x_{_\mu}\sqrt{x_{_\mu}x_{_1}}e^{i(\varphi_1+\varphi_\mu)}
{\cal C}_{_{131}}^0(x_{_\mu},x_{_1},x_{_{U_I}})
\nonumber\\&&\hspace{3.0cm} -4s_{_\beta}x_{_\mu}{\cal
C}_{_{131}}^1(x_{_\mu},x_{_1},x_{_{U_I}})
\Big]\bar{t}_{_R}{1\over/\!\!\!p_1+/\!\!\!q_1}q_1\cdot p_1t_{_R}
\nonumber\\&&\hspace{3.0cm}
+4x_{_{U_I}}\Big[c_{_\beta}x_{_\mu}\sqrt{x_{_\mu}x_{_1}}e^{i(\varphi_1+\varphi_\mu)}
{\cal C}_{_{113}}^0(x_{_\mu},x_{_1},x_{_{U_I}})
\nonumber\\&&\hspace{3.0cm} -s_{_\beta}{\cal
C}_{_{113}}^1(x_{_\mu},x_{_1},x_{_{U_I}})\Big]
\bar{t}_{_R}{1\over/\!\!\!p_1+/\!\!\!q_1}p_1\cdot(q_1+p_1)t_{_R}
\nonumber\\&&\hspace{3.0cm} -{\cal
C}_{_{112}}^1(x_{_\mu},x_{_1},x_{_{U_I}})
\bar{t}_{_R}{1\over/\!\!\!p_1+/\!\!\!q_1}[/\!\!\!q_1,/\!\!\!p_1]t_{_R}
\bigg\}, \label{example3}
\end{eqnarray}
where $\varphi_{\mu}$ and $\varphi_i$ (i=1,2,3) denote the $CP$-phases of
the parameter $\mu$ and $m_i$, respectively. As already mentioned before,
the first term of Eq. (\ref{example3}) is related to the Yukawa
coupling renormalization in the full theory and it does not affect
our computation. While matching Eq. (\ref{example3}) with
Eq. (\ref{example2}), we find that the diagram does not contribute
to the Wilson coefficients of the operators ${\cal O}_{_{t\Phi}}$.
A similar conclusion is true also for the other 1PI diagrams in
Fig. \ref{fig6}.

\subsection{The anomalous couplings ${\cal O}_{_{q\Phi}}$}

\indent\indent This class of anomalous couplings includes the effective operators
\begin{eqnarray}
&&{\cal O}_{_{q\Phi1}}=i\Big(\Phi^\dagger D_{_\mu}\Phi-(D_{_\mu}\Phi)^\dagger
\Phi\Big)\bar{q}_{_L}\gamma^\mu q_{_L}\;,\nonumber\\
&&{\cal O}_{_{q\Phi2}}=i\Big(\Phi^\dagger\sigma^aD_{_\mu}\Phi-(D_{_\mu}\Phi)^\dagger
\sigma^a\Phi\Big)\bar{q}_{_L}\sigma^a\gamma^\mu q_{_L}\;,\nonumber\\
&&{\cal O}_{_{q\Phi3}}=i\Big(\Phi^\dagger\Phi\Big)\Big(\bar{q}_{_L}
\gamma^\mu(D_{_\mu}q_{_L})-\overline{(D_{_\mu}q_{_L})}\gamma^\mu
q_{_L}\Big)\;,\nonumber\\
&&{\cal O}_{_{q\Phi4}}=i\Big(\Phi^\dagger\sigma^a\Phi\Big)\Big(\bar{q}_{_L}
\sigma^a\gamma^\mu(D_{_\mu}q_{_L})-\overline{(D_{_\mu}q_{_L})}
\sigma^a\gamma^\mu q_{_L}\Big)\;,\nonumber\\
&&{\cal O}_{_{q\Phi5}}=i\Big(\Phi^\dagger D_{_\mu}\Phi+(D_{_\mu}\Phi)^\dagger
\Phi\Big)\bar{q}_{_L}\gamma^\mu q_{_L}\;,\nonumber\\
&&{\cal O}_{_{q\Phi6}}=i\Big(\Phi^\dagger\sigma^aD_{_\mu}\Phi+(D_{_\mu}\Phi)^\dagger
\sigma^a\Phi\Big)\bar{q}_{_L}\sigma^a\gamma^\mu q_{_L}\;,\nonumber\\
\label{op3}
\end{eqnarray}
where the last two operators are $CP$-odd  and the others are
$CP$-even. The Feynman diagrams, which induce nontrivial
contributions to the Wilson coefficients, are presented in Fig.
\ref{fig8}. We collect the expressions for the Wilson coefficients
of the corresponding operators in Appendix \ref{aop3}.

\section{Experimental bounds on the Wilson coefficients}

\indent\indent
At present, the most rigorous constraint on
the Wilson coefficients considered in this work comes from the decay
$Z\rightarrow b\bar{b}$. For an on-shell $Z$, one can write the
general effective vertex $Zb\bar{b}$  as  \cite{Whisnant}
\begin{equation}
\Gamma_\mu^{Zbb}=-i{e\over4s_{_{\rm W}}c_{_{\rm W}}}\Big[V_b^Z\gamma_\mu
-A_b^Z\gamma_\mu\gamma_5+{1\over2m_{_b}}S^Z_b(p_{_b}-p_{_{\bar{b}}})\Big]
\;,
\label{verzbb}
\end{equation}
where $s_{_{\rm W}}\equiv \sin\theta_{_{\rm W}},\;c_{_{\rm
W}}\equiv \cos\theta_{_{\rm W}}$, and $p_{_b},\;p_{_{\bar{b}}}$
are the momenta of the outgoing quark and antiquark, respectively.
For the operators listed in Eqs. (\ref{op1e}), (\ref{op1o}),
(\ref{op2}) and (\ref{op3}), $S_b^Z=0$. The vector and
axial-vector couplings can be written as
\begin{eqnarray}
&&V_b^Z=V_b^{Z,0}+\delta V_b^Z\;,\nonumber\\
&&A_b^Z=A_b^{Z,0}+\delta A_b^Z\;,
\label{zbb1}
\end{eqnarray}
where $V_b^{Z,0},\;A_b^{Z,0}$ represent the SM couplings and $\delta
V_b^Z,\;\delta A_b^Z$ are the new physics contributions. Ignoring
the bottom quark mass, the lowest order theoretical prediction on
the observable $R_b$ at the $Z$ pole is given as
\begin{eqnarray}
&&R_b={\Gamma(Z\rightarrow b\bar{b})\over\Gamma(Z\rightarrow hadrons)}
=R_b^{SM}\bigg\{1+2{V_b^{Z,0}\delta V_b^{Z}+A_b^{Z,0}\delta A_b^Z\over
\Big(V_b^{Z,0}\Big)^2+\Big(A_b^{Z,0}\Big)^2}(1-R_b^{SM})\bigg\}\;.
\label{zbb2}
\end{eqnarray}

With the Born approximation,  we can obtain  modifications to the
couplings $V_b^Z\;,A_b^Z$ induced by the new physics operators
${\cal O}_{_{q\Phi1}}$ and ${\cal O}_{_{q\Phi2}}$. Provided that
there is no accidental cancellation between these contributions,
the corrections are  given by \cite{Lin}
\begin{equation}
\delta V_b^Z=\delta A_b^Z={2s_{_{\rm W}}m_{_{\rm W}}\upsilon\over e\mu_{_{NP}}^2}
\Big[C_{_{q\Phi1}}+C_{_{q\Phi2}}\Big]\;,
\label{zbb3}
\end{equation}
where $\upsilon$ denotes the VEV of the SM Higgs field doublet and
$m_{_{\rm W}}$ is the $W$-boson mass. From Eq. (\ref{zbb2}), we
have
\begin{eqnarray}
&&\delta V_b^Z=\delta A_b^Z={R_b^{exp}-R_b^{SM}\over(1-R_b^{SM})R_b^{SM}}
{\Big(V_b^{Z,0}\Big)^2+\Big(A_b^{Z,0}\Big)^2\over2(V_b^{Z,0}+A_b^{Z,0})}\;.
\label{zbb4}
\end{eqnarray}
The SM prediction on $R_b$ and the most recent experimental value
are, respectively, given by \cite{Erler}:
\begin{eqnarray}
&&R_b^{SM}=0.21572\pm0.00015,\;\;R_b^{exp}=0.21664\pm0.00065\;.
\label{zbb5}
\end{eqnarray}
If we attribute the difference of these two values to the new physics effects, we
get a bound for the new physics corrections on the $R_{_b}$. At the
$1\sigma$ tolerance we obtain:
\begin{equation}
0.00012\leq \Delta R_{_b}\leq 0.00172\;.
\label{zbb5-6}
\end{equation}
Correspondingly, the bound for the Wilson coefficients is
\begin{equation}
3.1\times 10^{-4}\leq
{\upsilon^2\over\mu_{_{NP}}^2}C_{_{q\Phi(1+2)}}\leq
4.5\times10^{-3} \label{zbb6}
\end{equation}
with $C_{_{q\Phi(1+2)}}=C_{_{q\Phi1}}+C_{_{q\Phi2}}$.
Using the same method, we can also analyze
the forward-backward asymmetry, $A_{FB}^b$, of the decay $Z\rightarrow b\bar{b}$.
However, our theoretical result indicates that the
present experiment data on this quantity set  a weaker bound on the Wilson
coefficients than $R_b$.

The other Wilson coefficients of the operators appearing in the
Lagrangian are not constrained by  $R_b$ on the Born
approximation level. With higher-order approximations, those
operators contribute to the gauge boson self-energies, and thus we
can get for them only a rather loose bound  with a significant
uncertainty. We can also have loose bounds from the argument of
partial wave unitarity \cite{Gounaris}:
\begin{eqnarray}
&&|C_{_{tq\Phi1}}|\le {16\pi\over3\sqrt{2}}\Big({\mu_{_{NP}}\over\upsilon}\Big)
\;,\;\;|C_{_{t\Phi1}}|\le 8\pi\sqrt{3}\;,\nonumber\\
&& -6.4\le C_{_{tq\Phi2}} \le 10.4\;.
\label{zbb7}
\end{eqnarray}
At present, there are no strong experimental constraints on the
$CP$-odd couplings involving the top quark.

It is well known that the MSSM contains in its general form unfortunately many 'new' free
parameters in addition to the SM parameters. In order to simplify our
discussion, we take the following assumption to restrict the MSSM
parameter space:
\begin{itemize}
\item All  possible $CP$ phases are taken to be zero or $\pi$.
A direct consequence of this choice is that there are no $CP$-odd
operators in the effective Lagrangian Eq. (\ref{elag})
 \item All
Yukawa couplings and the soft breaking parameters are flavor
conserving, {\it i.e.}, the mixing matrices ${\cal Z}_{_Q} ={\cal
Z}_{_U}={\cal Z}_{_D}={\bf 1}$.
\end{itemize}
Under these assumptions, the parameters relevant to our discussion
are the gauge coupling constants $g_1,\;g_2,\; g_3$, the higgsino
and gaugino masses $\mu,\;m_1,\;m_2,\;m_3$, the Yukawa couplings
of the third generation quarks and the corresponding soft breaking
parameters $h_t=h_{_U}^{33},\;h_b=h_{_D}^{33},\;
A_t=A_{_U}^{33},\;A_b=A_{_D}^{33}$, and the square masses of the
heavy Higgs boson doublet and the third generation squarks
$m_{_H}^2,\;m_{_{Q_I}}^2,\; m_{_{U_I}}^2,\;m_{_{D_I}}^2\;(I=3)$.
In our numerical analysis, we will disregard the loose bounds from
partial wave unitarity on the Wilson coefficients
$C_{_{tq\Phi1}},\; C_{_{t\Phi1}},\;C_{_{tq\Phi2}}$ due to the
large uncertainties mentioned above.

Without losing generality, we  assume $m_{_Q}=m_{_U}=m_{_D},\;
A_t=A_b,\;m_1=m_2=m_3$ in our numerical computations. Setting
$\mu_{_{NP}}=1000\;  {\rm GeV},\;m_{_H}=500\; {\rm
GeV},\;m_1=m_2=m_3 =500\; {\rm GeV},A_t=A_b=100\; {\rm GeV}$, we
obtain constraints set by Eq. (\ref{zbb6}) on the soft breaking
parameters. In Fig. \ref{fig9}, we plot the values of
$m_{_Q}=m_{_U}=m_{_D}$ versus the parameter $\mu$ with (a)
$\tan\beta=2$, and (b) $\tan\beta=40$, where the gray regions are
allowed by the condition for
${\upsilon^2C_{_{q\Phi(1+2)}}/\mu_{_{NP}}^2}$ set by Eq.
(\ref{zbb6}). From this plot we observe that the restriction set
on the parameter space with $\tan\beta=40$ is more rigorous than
that with $\tan\beta=2$. As $\tan\beta=2$, the contribution from
the supersymmetric box diagrams varies from negative to positive
gradually, then tends to zero after its maximum as
$m_{_Q}=m_{_U}=m_{_D}$ increase from $200\;{\rm GeV}$. When
$m_{_Q}=m_{_U}=m_{_D}\ge 1.3\;{\rm TeV}$, the contribution is
definitely less than $10^{-4}$. Beside  those box diagrams,
$C_{_{q\Phi1}}$ also receives a contribution from the heavy Higgs
doublet. Under our choice about the parameter space, the Higgs
contribution to the Wilson coefficient
$\upsilon^2C_{_{q\Phi1}}/\mu_{_{\rm NP}}^2$ is proportional to
$\Big[1/\tan^2\beta-\Big(m_{_b}/ m_{_t}\Big)^4\tan^2 \beta\Big]$.
Taking the bottom quark mass $m_{_b}=4.5\;{\rm GeV}$ and top mass
$m_{_t}=174\;{\rm GeV}$, this contribution is about
$5\times10^{-4}$ for $\tan\beta=2$. As $\tan\beta$ increasing, the
contribution of the heavy Higgs doublet is  strongly suppressed
and less than $10^{-7}$ for $\tan\beta=40$. The fact can help us
understanding why very massive supersymmetry particles are allowed
by the experimental bound  for $\tan\beta=2$ (Fig. \ref{fig9}(a)),
whereas the most part of the parameter space  is excluded by the
bound except a narrow band at the neighborhood of $\mu=0$   for
$\tan\beta=40$ (Fig. \ref{fig9}(b)).

In the figures discussed above, we have considered the $1\sigma$
tolerance for the experimental data.
Since the central value of the experimentally measured $R_b$ is
only about one standard deviation away from the SM prediction,
this sets a lower bound on the $C_{_{q\Phi(1+2)}}$  which is
positive and very close to zero as shown in Eq.(\ref{zbb6}).
Certainly, very massive supersymmetry particles are excluded by
this condition in the large $\tan\beta$ case. In fact, considering
the practical situation of
the experiments, we may relax the lower bound on the
$C_{_{q\Phi(1+2)}}$ to $-5\times10^{-5}$, while the upper bound
remains unchanged (this is just only slightly
beyond the standard deviation). In Fig. \ref{fig10} we plot
$m_{_Q}=m_{_U}=m_{_D}$ versus $\mu$ by using the constraint
$-5\times 10^{-5}\le {\upsilon^2C_{_{q\Phi(1+2)}}/\mu_{_{NP}}^2}\le
4.5\times10^{-3}$. One can see that the allowed parameter region
is drastically enlarged in comparison with the case of the strict
$1\sigma$ tolerance for the large $\tan\beta$.

Now, we discuss the operators ${\cal O}_{_{q\Phi1,2}}$ corrections
to $R_{_b}$ in the MSSM. Taking $\mu=m_i=500\;{\rm
GeV}\;(i=1,\;2,\;3)$, we plot $\Delta R_{_b}$ versus squark masses
$m_{_Q}=m_{_U}=m_{_D}$ with $\tan\beta=2,\;40$ in Fig.
\ref{fig11}. The gray band is the experimentally allowed region at
the $1\sigma$ tolerance. When the scalar quark mass is less than
$700\;{\rm GeV}$,  the supersymmetric box diagrams determine the
leading contribution and results in a negative $\Delta R_{_b}$,
the corresponding parameter space is excluded by Eq.
(\ref{zbb5-6}). As the parameters $m_{_Q}=m_{_U}=m_{_D}$ increase,
the supersymmetric contribution turns to be positive, then tends
to zero after the maximum. With $\tan\beta=2$, the correction of
the heavy Higgs doublet to the $\Delta R_{_b}$ is about $2.3\times
10^{-4}$, and plays the leading role when
$m_{_Q}=m_{_U}=m_{_D}\ge1.3\;{\rm TeV}$. For $\tan\beta=40$, the
total corrections from the Higgs and supersymmetric sectors to
$R_{_b}$ do not satisfy Eq. (\ref{zbb5-6}), because the
contribution of the heavy Higgs doublet  is  strongly suppressed.
For $-\mu=m_i=500\;{\rm GeV}\;(i=1,\;2,\;3)$, the plot is similar
to Fig. \ref{fig11} and not shown in the context. Taking
$m_i=500\;{\rm GeV}\;(i=1,\;2,\;3)$,
$m_{_Q}=m_{_U}=m_{_D}=500,\;1000\;{\rm GeV}$, and
$\tan\beta=2,\;40$, we present $\Delta R_{_b}$ versus the
parameter $\mu$ in Fig. \ref{fig12}. For
$m_{_Q}=m_{_U}=m_{_D}=500\;{\rm GeV}$ (dot- and dot-dashed-lines),
the corresponding parameter space is excluded by the condition Eq.
(\ref{zbb5-6}) due to the negative supersymmetry contribution.
With $m_{_Q}=m_{_U}=m_{_D}=1\;{\rm TeV}$ and $\tan\beta=2$
(solid-line), $\Delta R_{_b}$ satisfies the condition Eq.
(\ref{zbb5-6}) when $\mu\ge-700\;{\rm GeV}$. As for the case
$m_{_Q}=m_{_U}=m_{_D}=1\;{\rm TeV}$ and $\tan\beta=40$
(dashed-line), the new physics correction to $\Delta R_{_b}$ is
excluded by the 1$\sigma$ tolerance experimental bound except the
region neighboring $\mu=0\;{\rm GeV}$. Finally, we investigate the
new physics prediction on $\Delta R_{_b}$ with the assumption
$m_{_Q}=m_{_U}=m_{_D}=|\mu|$. Choosing $m_i=500\;{\rm
GeV}\;(i=1,\;2,\;3)$, $\tan\beta=2,\;40$, we plot $\Delta R_{_b}$
versus the parameter $m_{_Q}=m_{_U}=m_{_D}=|\mu|$ in Fig.
\ref{fig13}. For the case $\tan\beta=40$, the correction to
$\Delta R_{_b}$ exceeds the 1$\sigma$ tolerance experimental
bound. As $\tan\beta=2$ and $\mu>800\;{\rm GeV}$, the new physics
prediction on $\Delta R_{_b}$ satisfies this bound because of the
relatively large contribution from the heavy Higgs. In those
analyses the experimental bound with $1\sigma$ standard deviation
are adopted. After we relax the condition (Eq. \ref{zbb5-6})
slightly, the more massive supersymmetry particles are also
permitted by the corresponding experimental bound.

Since the experimental data constrain the coefficients
$C_{_{q\Phi1,2}}$ strongly, the operators ${\cal O}_{_{q\Phi1,2}}$
have only negligible effects on the measurements at the
proposed future colliders \cite{Whisnant}. Other operators will
produce the observable effects in the next generation colliders.
In the associated production of the Higgs boson and top quark pair
$e^+e^-\rightarrow t\bar{t}\;h$, the $CP$-even operators will
affect the energy and angular distributions of the final state
particles \cite{Lin}. Through the measurements of various
distributions, such as ${d\sigma/dE_{_t}},\;{d\sigma/
dE_{_h}}$ and ${d\sigma/d\cos\theta_{_h}}$ ($E_{_t},\;E_{_h}$
denote the outgoing energy of the top quark and Higgs boson
respectively, $\theta_{_h}$ is the angle of three-momentum of the
outgoing Higgs boson with respect to the electron beam direction),
we can obtain useful information about the operators. The
constraints on the $CP$-odd operators can be obtained through
measuring various $CP$ violation observables in this process. In the process
$e^+e^-\rightarrow t\bar{t}$, we can analyze the effects of the
operators on various polarized top-quark production cross
sections. On the other hand, more strict constraints on the
supersymmetry parameter space will be set by more precise
measurements on the widths of $Z\rightarrow b\bar{b}$ and the top quark
decays. All of these will provide valuable information for the search of
supersymmetry particles on the future colliders.

It should be stressed that the above numerical analysis is
performed under special assumptions about the MSSM parameter
space.  For example, we assume that all the parameters are real
and flavor-conserving, the universal soft parameters are:
$m_{_Q}=m_{_U}=m_{_D},\;A_t=A_b,\;m_1=m_2=m_3$. In a practical
phenomenology analysis, those priori conditions should be
dismissed. Nevertheless, it is quite obvious that the experimental
data on $R_b$ set significant bounds on the parameter space even
in a more general case than that we have considered here.

\section{The Summary}

\indent\indent We have considered in this work the anomalous
couplings between  top quark and  Higgs boson induced by the MSSM
when the heavy Higgs doublet and all supersymmetry fields are
integrated out. An essential assumption made here is that  there
is only one neutral Higgs boson with the electroweak mass, the
other Higgs particles are much heavier. We have derived the Wilson
coefficients of the relevant higher dimensional operators in the
ensuing effective theory. We have also studied numerically the
constraints set by the experimental results for
$R_b={\Gamma(Z\rightarrow b\bar{b})/\Gamma(Z\rightarrow hadrons)}$
on the parameters of the MSSM.

\vspace{1cm} \noindent{\bf Acknowledgement}

The work has been supported by the Academy of Finland under the contract
no.\ 40677 and by the National Natural Science Foundation of China.

\vspace{2cm}

\appendix

\section{The Wilson coefficients for the operators ${\cal O}_{_{tq\Phi i}}
\;(i=2,\;\cdots,5,\;7,\;\cdots,\;10)$ \label{aop1}}

\begin{eqnarray}
&&C_{_{tq\Phi2}}=
-{c_{_\beta}\over(4\pi)^2}\bigg\{\sum\limits_{i=1,2}
F_ig_i^2\xi_{_{Hi}}^RP_{_H}(x_{_\mu},x_{_i})
-s_{_\beta}c_{_\beta}\sum\limits_{q=U,D}\Big(-1\Big)^{{1\over2}-T_Z^q}
{\bf Re}\Big[{\bf h}_{_U}^{33}\Big({\cal Z}_{_q}\hat{\bf A}_{_q}{\cal Z}_{_Q}\Big)_{IJ}
\nonumber\\
&&\hspace{1.5cm}\times
\Big({\cal Z}_{_q}\hat{\bf A}_{_q}^\prime{\cal Z}_{_Q}\Big)^\dagger_{JI}\Big]
P_{_H}(x_{_{Q_J}},x_{_{q_I}})+s_{_\beta}c_{_\beta}
{\bf Re}\Big[{\bf h}_{_U}^{33}\Big({\cal Z}_{_R}\hat{\bf A}_{_E}{\cal Z}_{_L}\Big)_{IJ}
\Big({\cal Z}_{_R}\hat{\bf A}_{_E}^\prime{\cal Z}_{_L}\Big)^\dagger_{JI}\Big]
P_{_H}(x_{_{L_J}},x_{_{R_I}})\bigg\}
\nonumber\\
&&\hspace{1.5cm}
+{1\over32\pi^2x_{_H}}s_{_\beta}c_{_\beta}^2{\bf Re}\Big({\bf h}_{_D}^\dagger
{\bf h}_{_D}{\bf h}_{_U}^\dagger\Big)_{33}\Big(1-\ln x_{_H}\Big)
+{g_3^2\over6\pi^2}s_{_\beta}\Gamma_{_{U,IJ}}^{^{R,3}}
x_{_{U_J}}{\cal C}_{_{113}}^0(x_{_3},x_{_{Q_I}},x_{_{U_J}})
\nonumber\\
&&\hspace{1.5cm}
+{g_1^2\over72\pi^2}s_{_\beta}\Gamma_{_{U,IJ}}^{^{R,1}}
x_{_{U_J}}{\cal C}_{_{113}}^0(x_{_1},x_{_{Q_I}},x_{_{U_J}})
+{1\over16\pi^2}c_{_\beta}\Gamma_{_{U,IJ}}^{^S}
x_{_{Q_J}}{\cal C}_{_{113}}^0(x_{_\mu},x_{_{D_I}},x_{_{Q_J}})
\nonumber\\
&&\hspace{1.5cm}
-{1\over192\pi^2}\sum\limits_{i=1,2}F_i^2g_i^2\Big[\Lambda_{_{Q,I}}^{^T}
\Big({\cal C}_{_{211}}^1(x_{_\mu},x_{_i},x_{_{Q_I}})
+2s_{_\beta}x_{_\mu}{\cal C}_{_{311}}^1(x_{_\mu},x_{_i},x_{_{Q_I}})\Big)
\nonumber\\
&&\hspace{1.5cm}
+2c_{_\beta}x_{_\mu}\Lambda_{_{U,I}}^{^{R,i}}
{\cal C}_{_{311}}^0(x_{_\mu},x_{_i},x_{_{Q_I}})\Big]
-{1\over48\pi^2}g_1^2\Big[\Lambda_{_{U,I}}^{^T}
\Big({\cal C}_{_{121}}^1(x_{_\mu},x_{_1},x_{_{U_I}})
\nonumber\\
&&\hspace{1.5cm}
+2s_{_\beta}x_{_\mu}{\cal C}_{_{131}}^1(x_{_\mu},x_{_1},x_{_{U_I}})\Big)
+2c_{_\beta}x_{_\mu}\Lambda_{_{U,I}}^{^{R,1}}
{\cal C}_{_{131}}^0(x_{_\mu},x_{_1},x_{_{U_I}})\Big]
\;,\nonumber\\
&&C_{_{tq\Phi3}}=
-{1\over32\pi^2x_{_H}}s_{_\beta}c_{_\beta}^2{\bf Re}\Big({\bf h}_{_D}^\dagger
{\bf h}_{_D}{\bf h}_{_U}^\dagger\Big)_{33}
+{g_3^2\over6\pi^2}s_{_\beta}\Gamma_{_{U,IJ}}^{^{R,3}}
Q_1(x_{_3},x_{_{Q_I}},x_{_{U_J}})
\nonumber\\
&&\hspace{1.5cm}
+{g_1^2\over72\pi^2}s_{_\beta}
\Gamma_{_{U,IJ}}^{^{R,1}}Q_1(x_{_1},x_{_{Q_I}},x_{_{U_J}})
+{1\over16\pi^2}c_{_\beta}\Gamma_{_{U,IJ}}^{^S}
Q_1(x_{_\mu},x_{_{D_I}},x_{_{Q_J}})
\nonumber\\
&&\hspace{1.5cm}
-{1\over96\pi^2}\sum\limits_{i=1,2}F_i^2g_i^2x_{_{Q_I}}
\Big[s_{_\beta}\Lambda_{_{Q,I}}^{^T}
{\cal C}_{_{113}}^1(x_{_\mu},x_{_i},x_{_{Q_I}})
+c_{_\beta}x_{_\mu}\Lambda_{_{U,I}}^{^{R,i}}
{\cal C}_{_{113}}^0(x_{_\mu},x_{_i},x_{_{Q_I}})\Big]
\nonumber\\
&&\hspace{1.5cm}
-{1\over24\pi^2}g_1^2x_{_{U_I}}
\Big[s_{_\beta}\Lambda_{_{U,I}}^{^T}
{\cal C}_{_{113}}^1(x_{_\mu},x_{_1},x_{_{U_I}})
+c_{_\beta}x_{_\mu}\Lambda_{_{U,I}}^{^{R,1}}
{\cal C}_{_{113}}^0(x_{_\mu},x_{_1},x_{_{U_I}})\Big]
\;,\nonumber\\
&&C_{_{tq\Phi4}}=
-{c_{_\beta}\over(4\pi)^2}\bigg\{\sum\limits_{i=1,2}
F_ig_i^2\xi_{_{Hi}}^RP_{_H}(x_{_\mu},x_{_i})
-s_{_\beta}c_{_\beta}\sum\limits_{q=U,D}\Big(-1\Big)^{{1\over2}-T_Z^q}
{\bf Re}\Big[{\bf h}_{_U}^{33}\Big({\cal Z}_{_q}\hat{\bf A}_{_q}{\cal Z}_{_Q}\Big)_{IJ}
\nonumber\\
&&\hspace{1.5cm}\times
\Big({\cal Z}_{_q}\hat{\bf A}_{_q}^\prime{\cal Z}_{_Q}\Big)^\dagger_{JI}\Big]
P_{_H}(x_{_{Q_J}},x_{_{q_I}})+s_{_\beta}c_{_\beta}
{\bf Re}\Big[{\bf h}_{_U}^{33}\Big({\cal Z}_{_R}\hat{\bf A}_{_E}{\cal Z}_{_L}\Big)_{IJ}
\Big({\cal Z}_{_R}\hat{\bf A}_{_E}^\prime{\cal Z}_{_L}\Big)^\dagger_{JI}\Big]
P_{_H}(x_{_{L_J}},x_{_{R_I}})\bigg\}
\nonumber\\
&&\hspace{1.5cm}
+{1\over32\pi^2x_{_H}}s_{_\beta}c_{_\beta}^2{\bf Re}\Big({\bf h}_{_D}^\dagger
{\bf h}_{_D}{\bf h}_{_U}^\dagger\Big)_{33}\Big(1-\ln x_{_H}\Big)
+{g_3^2\over6\pi^2}s_{_\beta}\Gamma_{_{U,IJ}}^{R,3}
x_{_{Q_I}}{\cal C}_{_{131}}^0(x_{_3},x_{_{Q_I}},x_{_{U_J}})
\nonumber\\
&&\hspace{1.5cm}
+{g_1^2\over72\pi^2}s_{_\beta}\Gamma_{_{U,IJ}}^{R,1}
x_{_{Q_I}}{\cal C}_{_{131}}^0(x_{_1},x_{_{Q_I}},x_{_{U_J}})
+{1\over16\pi^2}c_{_\beta}\Gamma_{_{U,IJ}}^{^S}
x_{_{D_I}}{\cal C}_{_{131}}^0(x_{_\mu},x_{_{D_I}},x_{_{Q_J}})
\nonumber\\
&&\hspace{1.5cm}
-{1\over192\pi^2}\sum\limits_{i=1,2}F_i^2g_i^2
\Big[\Lambda_{_{Q,I}}^{^T}\Big(Q_9(x_{_\mu},x_{_i},x_{_{Q_I}})
+2s_{_\beta}Q_2(x_{_i},x_{_\mu},x_{_{Q_I}})\Big)
\nonumber\\
&&\hspace{1.5cm}
+2c_{_\beta}\Lambda_{_{U,I}}^{^{R,i}}
Q_1(x_{_i},x_{_\mu},x_{_{Q_I}})\Big]
-{1\over48\pi^2}g_1^2\Big[\Lambda_{_{U,I}}^{^T}
\Big(Q_9(x_{_1},x_{_\mu},x_{_{U_I}})
\nonumber\\
&&\hspace{1.5cm}
+2s_{_\beta}Q_4(x_{_\mu},x_{_1},x_{_{U_I}})\Big)
+2c_{_\beta}\Lambda_{_{U,I}}^{^{R,1}}
Q_3(x_{_\mu},x_{_1},x_{_{Q_I}})\Big]
\;,\nonumber\\
&&C_{_{tq\Phi5}}=
-{1\over32\pi^2x_{_H}}s_{_\beta}c_{_\beta}^2{\bf Re}\Big({\bf h}_{_D}^\dagger
{\bf h}_{_D}{\bf h}_{_U}^\dagger\Big)_{33}
-{1\over192\pi^2}\sum\limits_{i=1,2}F_i^2g_i^2\Lambda_{_{Q,I}}^{^T}
{\cal C}_{_{112}}^1(x_{_\mu},x_{_i},x_{_{Q_I}})
\nonumber\\
&&\hspace{1.5cm}
-{1\over48\pi^2}g_1^2\Lambda_{_{U,I}}^{^T}
{\cal C}_{_{112}}^1(x_{_\mu},x_{_1},x_{_{U_I}})
\;,\nonumber\\
&&C_{_{tq\Phi7}}=
i{c_{_\beta}\over(4\pi)^2}\bigg\{\sum\limits_{i=1,2}F_ig_i^2\xi_{_{Hi}}^A
P_{_H}(x_{_\mu},x_{_i})-s_{_\beta}c_{_\beta}\sum\limits_{q=U,D}\Big(-1\Big)^{{1\over2}-T_Z^q}
{\bf Im}\Big[{\bf h}_{_U}^{33}\Big({\cal Z}_{_q}\hat{\bf A}_{_q}{\cal Z}_{_Q}\Big)_{IJ}
\nonumber\\
&&\hspace{1.5cm}\times
\Big({\cal Z}_{_q}\hat{\bf A}_{_q}^\prime{\cal Z}_{_Q}\Big)^\dagger_{JI}\Big]
P_{_H}(x_{_{Q_J}},x_{_{q_I}})+s_{_\beta}c_{_\beta}
{\bf Im}\Big[{\bf h}_{_U}^{33}\Big({\cal Z}_{_R}\hat{\bf A}_{_E}{\cal Z}_{_L}\Big)_{IJ}
\Big({\cal Z}_{_R}\hat{\bf A}_{_E}^\prime{\cal Z}_{_L}\Big)^\dagger_{JI}\Big]
P_{_H}(x_{_{L_J}},x_{_{R_I}})\bigg\}
\nonumber\\
&&\hspace{1.5cm}
+{i\over32\pi^2x_{_H}}s_{_\beta}c_{_\beta}^2{\bf Im}\Big({\bf h}_{_D}^\dagger
{\bf h}_{_D}{\bf h}_{_U}^\dagger\Big)_{33}\Big(1-\ln x_{_H}\Big)
+i{g_3^2\over6\pi^2}s_{_\beta}\Gamma_{_{U,IJ}}^{^{A,3}}
x_{_{U_J}}{\cal C}_{_{113}}^0(x_{_3},x_{_{Q_I}},x_{_{U_J}})
\nonumber\\
&&\hspace{1.5cm}
+i{g_1^2\over72\pi^2}s_{_\beta}\Gamma_{_{U,IJ}}^{^{A,1}}
x_{_{U_J}}{\cal C}_{_{113}}^0(x_{_1},x_{_{Q_I}},x_{_{U_J}})
+{i\over16\pi^2}c_{_\beta}\Gamma_{_{U,IJ}}^{^B}
x_{_{Q_J}}{\cal C}_{_{113}}^0(x_{_\mu},x_{_{D_I}},x_{_{Q_J}})
\nonumber\\
&&\hspace{1.5cm}
-i{1\over192\pi^2}\sum\limits_{i=1,2}F_i^2g_i^2\Big[\Lambda_{_{Q,I}}^{^C}
\Big({\cal C}_{_{211}}^1(x_{_\mu},x_{_i},x_{_{Q_I}})
+2s_{_\beta}x_{_\mu}{\cal C}_{_{311}}^1(x_{_\mu},x_{_i},x_{_{Q_I}})\Big)
\nonumber\\
&&\hspace{1.5cm}
+2c_{_\beta}x_{_\mu}\Lambda_{_{U,I}}^{^{A,i}}
{\cal C}_{_{311}}^0(x_{_\mu},x_{_i},x_{_{Q_I}})\Big]
-{i\over48\pi^2}g_1^2\Big[\Lambda_{_{U,I}}^{^C}
\Big({\cal C}_{_{121}}^1(x_{_\mu},x_{_1},x_{_{U_I}})
\nonumber\\
&&\hspace{1.5cm}
+2s_{_\beta}x_{_\mu}{\cal C}_{_{131}}^1(x_{_\mu},x_{_1},x_{_{U_I}})\Big)
+2c_{_\beta}x_{_\mu}\Lambda_{_{U,I}}^{^{A,1}}
{\cal C}_{_{131}}^0(x_{_\mu},x_{_1},x_{_{U_I}})\Big]
\;,\nonumber\\
&&C_{_{tq\Phi8}}=
-{i\over32\pi^2x_{_H}}s_{_\beta}c_{_\beta}^2{\bf Im}\Big({\bf h}_{_D}^\dagger
{\bf h}_{_D}{\bf h}_{_U}^\dagger\Big)_{33}
+i{g_3^2\over6\pi^2}s_{_\beta}\Gamma_{_{U,IJ}}^{^{A,3}}
Q_1(x_{_3},x_{_{Q_I}},x_{_{U_J}})
\nonumber\\
&&\hspace{1.5cm}
+i{g_1^2\over72\pi^2}s_{_\beta}\Gamma_{_{U,IJ}}^{^{A,1}}
Q_1(x_{_1},x_{_{Q_I}},x_{_{U_J}})
+{i\over16\pi^2}c_{_\beta}\Gamma_{_{U,IJ}}^{^B}
Q_1(x_{_\mu},x_{_{D_I}},x_{_{Q_J}})
\nonumber\\
&&\hspace{1.5cm}
-{i\over96\pi^2}\sum\limits_{i=1,2}F_i^2g_i^2x_{_{Q_I}}
\Big[s_{_\beta}\Lambda_{_{Q,I}}^{^C}
{\cal C}_{_{113}}^1(x_{_\mu},x_{_i},x_{_{Q_I}})
+c_{_\beta}x_{_\mu}\Lambda_{_{U,I}}^{^{A,i}}
{\cal C}_{_{113}}^0(x_{_\mu},x_{_i},x_{_{Q_I}})\Big]
\nonumber\\
&&\hspace{1.5cm}
-{i\over24\pi^2}g_1^2x_{_{U_I}}
\Big[s_{_\beta}\Lambda_{_{U,I}}^{^C}
{\cal C}_{_{113}}^1(x_{_\mu},x_{_1},x_{_{U_I}})
+c_{_\beta}x_{_\mu}\Lambda_{_{U,I}}^{^{A,1}}
{\cal C}_{_{113}}^0(x_{_\mu},x_{_1},x_{_{U_I}})\Big]
\;,\nonumber\\
&&C_{_{tq\Phi9}}=i{c_{_\beta}\over(4\pi)^2}\bigg\{\sum\limits_{i=1,2}F_ig_i^2\xi_{_{Hi}}^A
P_{_H}(x_{_\mu},x_{_i})-s_{_\beta}c_{_\beta}\sum\limits_{q=U,D}\Big(-1\Big)^{{1\over2}-T_Z^q}
{\bf Im}\Big[{\bf h}_{_U}^{33}\Big({\cal Z}_{_q}\hat{\bf A}_{_q}{\cal Z}_{_Q}\Big)_{IJ}
\nonumber\\
&&\hspace{1.5cm}\times
\Big({\cal Z}_{_q}\hat{\bf A}_{_q}^\prime{\cal Z}_{_Q}\Big)^\dagger_{JI}\Big]
P_{_H}(x_{_{Q_J}},x_{_{q_I}})+s_{_\beta}c_{_\beta}
{\bf Im}\Big[{\bf h}_{_U}^{33}\Big({\cal Z}_{_R}\hat{\bf A}_{_E}{\cal Z}_{_L}\Big)_{IJ}
\Big({\cal Z}_{_R}\hat{\bf A}_{_E}^\prime{\cal Z}_{_L}\Big)^\dagger_{JI}\Big]
P_{_H}(x_{_{L_J}},x_{_{R_I}})\bigg\}
\nonumber\\
&&\hspace{1.5cm}
+{i\over32\pi^2x_{_H}}s_{_\beta}c_{_\beta}^2{\bf Im}\Big({\bf h}_{_D}^\dagger
{\bf h}_{_D}{\bf h}_{_U}^\dagger\Big)_{33}\Big(1-\ln x_{_H}\Big)
+i{g_3^2\over6\pi^2}s_{_\beta}\Gamma_{_{U,IJ}}^{^{A,3}}
x_{_{Q_I}}{\cal C}_{_{131}}^0(x_{_3},x_{_{Q_I}},x_{_{U_J}})
\nonumber\\
&&\hspace{1.5cm}
+i{g_1^2\over72\pi^2}s_{_\beta}\Gamma_{_{U,IJ}}^{^{A,1}}
x_{_{Q_I}}{\cal C}_{_{131}}^0(x_{_1},x_{_{Q_I}},x_{_{U_J}})
+{i\over16\pi^2}c_{_\beta}\Gamma_{_{U,IJ}}^{^B}
x_{_{D_I}}{\cal C}_{_{131}}^0(x_{_\mu},x_{_{D_I}},x_{_{Q_J}})
\nonumber\\
&&\hspace{1.5cm}
-{i\over192\pi^2}\sum\limits_{i=1,2}F_i^2g_i^2
\Big[\Lambda_{_{Q,I}}^{^C}\Big(Q_9(x_{_\mu},x_{_i},x_{_{Q_I}})
+2s_{_\beta}Q_2(x_{_i},x_{_\mu},x_{_{Q_I}})\Big)
+2c_{_\beta}\Lambda_{_{U,I}}^{^{A,i}}
Q_1(x_{_i},x_{_\mu},x_{_{Q_I}})\Big]
\nonumber\\
&&\hspace{1.5cm}
-{i\over48\pi^2}g_1^2\Big[\Lambda_{_{U,I}}^{^C}\Big(Q_9(x_{_1},x_{_\mu},x_{_{U_I}})
+2s_{_\beta}Q_4(x_{_\mu},x_{_1},x_{_{U_I}})\Big)
+2c_{_\beta}\Lambda_{_{U,I}}^{^{A,1}}
Q_3(x_{_\mu},x_{_1},x_{_{Q_I}})\Big]
\;,\nonumber\\
&&C_{_{tq\Phi10}}=
-{i\over32\pi^2x_{_H}}s_{_\beta}c_{_\beta}^2{\bf Im}\Big({\bf h}_{_D}^\dagger
{\bf h}_{_D}{\bf h}_{_U}^\dagger\Big)_{33}
-{i\over192\pi^2}\sum\limits_{i=1,2}F_i^2g_i^2\Lambda_{_{Q,I}}^{^C}
{\cal C}_{_{112}}^1(x_{_\mu},x_{_i},x_{_{Q_I}})
\nonumber\\
&&\hspace{1.5cm}
-{i\over48\pi^2}g_1^2\Lambda_{_{U,I}}^{^C}
{\cal C}_{_{112}}^1(x_{_\mu},x_{_1},x_{_{U_I}})
\label{wcoe2}
\end{eqnarray}
with $F_1=1,\;F_2=3,\;T_Z^U=-T_Z^D={1\over2}$, $Y_{_U}^B={4\over3},
\;Y_{_D}^B=-{2\over3}$ and $\hat{\bf A}_{_{U}}={1\over\mu_{_{NP}}}
\Big({\bf A}_{_{U}}-{\mu^*\over\tan\beta}{\bf h}_{_U}\Big),\;
\hat{\bf A}_{_{D}}={1\over\mu_{_{NP}}}
\Big({\bf A}_{_{D}}-\mu^*\tan\beta{\bf h}_{_D}\Big),\;
\hat{\bf A}_{_{E}}={1\over\mu_{_{NP}}}
\Big({\bf A}_{_{E}}-\mu^*\tan\beta{\bf h}_{_E}\Big)$ and
$\hat{\bf A}_{_{F}}^\prime=\hat{\bf A}_{_{F}}+{\mu^*\over
s_{_\beta}c_{_\beta}\mu_{_{NP}}}{\bf h}_{_F}\;\;(F=U,\;D,\;E)$.
In the above expression, the sum with the generation
indices $I,\;J$ is implied.


\section{The Wilson coefficients for operators ${\cal O}_{_{t\Phi i}}\;
(i=1,\;2,\;3)$\label{aop2}}

The corresponding Wilson coefficients are
\begin{eqnarray}
&&C_{_{t\Phi1}}=-{1\over32\pi^2x_{_{H}}}c_{_\beta}^2\Big[{\bf h}_{_U}
\Big(s_{_\beta}^2{\bf h}_{_U}^\dagger{\bf h}_{_U}
-c_{_\beta}^2{\bf h}_{_D}^\dagger{\bf h}_{_D}\Big){\bf h}_{_U}^\dagger
\Big]_{33}\Big(1-\ln x_{_{H}}\Big)
\nonumber\\
&&\hspace{1.3cm}
+{{\cal X}_{_{U,IJK}}^{^R}\over72\pi^2}\Big[
g_1^2{\cal D}_{_{1121}}^1(x_{_1},
x_{_{U_I}},x_{_{Q_J}},x_{_{U_K}})
+3g_3^2{\cal D}_{_{1121}}^1(x_{_3},
x_{_{U_I}},x_{_{Q_J}},x_{_{U_K}})\Big]
\nonumber\\
&&\hspace{1.3cm}
+{1\over64\pi^2}\sum\limits_{q=U,D}{\cal X}_{_{q,IJK}}^{^S}
{\cal D}_{_{1121}}^1(x_{_\mu},x_{_{Q_I}},x_{_{q_J}},x_{_{Q_K}})
+{1\over144\pi^2}g_1^4{\cal Z}_{_U}^{\dagger3I}{\cal Z}_{_U}^{I3}
\Big[c_{_\beta}^2Q_5(x_{_\mu},x_{_1},x_{_{U_I}})
\nonumber\\
&&\hspace{1.3cm}
+2\xi_{_{H1}}^SQ_6(x_{_\mu},x_{_1},x_{_{U_I}})
+s_{_\beta}^2x_{_1}Q_7(x_{_\mu},x_{_1},x_{_{U_I}})\Big]
\nonumber\\
&&\hspace{1.3cm}
-\sum\limits_{i=1,2}{F_ig_i^2\over128\pi^2}\Big({\bf h}_{_U}{\cal Z}_{_Q}\Big)_{3I}
\Big({\cal Z}_{_Q}^\dagger {\bf h}_{_U}^\dagger\Big)_{I3}\Big[
2\xi_{_{Hi}}^SQ_6(x_{_i},x_{_\mu},x_{_{Q_I}})
\nonumber\\
&&\hspace{1.3cm}
+s_{_\beta}^2Q_5(x_{_i},x_{_\mu},x_{_{Q_I}})
+c_{_\beta}^2x_{_\mu}Q_7(x_{_i},x_{_\mu},x_{_{Q_I}})\Big]
\nonumber\\
&&\hspace{1.3cm}
+{1\over48\pi^2}g_1^2s_{_\beta}
\Big[2\Gamma_{_{U,IJ}}^{^U}{\cal D}_{_{1111}}^0(x_{_\mu},x_{_1},x_{_{Q_I}},
x_{_{U_J}})-\Gamma_{_{U,IJ}}^{^T}U_5(x_{_\mu},x_{_{Q_I}},x_{_1},
x_{_{U_J}})\Big]
\;,\nonumber\\
&&C_{_{t\Phi2}}={\Big({\bf h}_{_U}{\bf h}_{_U}^\dagger\Big)_{33}\over256\pi^2x_{_H}}
c_{_\beta}^2\Big[4s_{_\beta}^2(s_{_\beta}^2-c_{_\beta}^2)(2g_1^2+3g_2^2)\ln x_{_H}
-12s_{_\beta}^4g_2^2+(c_{_\beta}^4-7s_{_\beta}^4)g_1^2\Big]
\nonumber\\
&&\hspace{1.3cm}
+{g_1^2\over108\pi^2}(s_{_\beta}^2-c_{_\beta}^2){\cal Z}_{_U}^{\dagger 3I}
{\cal Z}_{_U}^{I3}\Big[g_1^2{\cal B}_{_{3,1}}^1(x_{_{U_I}},x_{_1})
+3g_3^2{\cal B}_{_{3,1}}^1(x_{_{U_I}},x_{_3})\Big]
\nonumber\\
&&\hspace{1.3cm}
+{c_{_\beta}^2\over72\pi^2}{\bf Re}\Big[{\cal Z}_{_U}^{\dagger3I}
\Big({\cal Z}_{_U}{\bf h}_{_U}{\bf h}_{_U}^\dagger{\cal Z}_{_U}^\dagger
\Big)^{IJ}{\cal Z}_{_U}^{J3}\Big]\Big[g_1^2Q_9(x_{_{U_I}},x_{_1},x_{_{U_J}})
+3g_3^2Q_9(x_{_{U_I}},x_{_3},x_{_{U_J}})\Big]
\nonumber\\
&&\hspace{1.3cm}
-{1\over192\pi^2}g_1^2(s_{_\beta}^2-c_{_\beta}^2)\Big(
{\bf h}_{_U}{\cal Z}_{_Q}\Big)_{3I}\Big({\cal Z}_{_Q}^\dagger
{\bf h}_{_U}^\dagger\Big)_{I3}{\cal B}_{_{3,1}}^1(x_{_{Q_I}},x_{_\mu})
\nonumber\\
&&\hspace{1.3cm}
+{1\over16\pi^2}{\bf Re}\Big[\Big({\bf h}_{_U}{\cal Z}_{_Q}\Big)^{3I}
\Big({\cal Z}_{_Q}^\dagger(c_{_\beta}^2{\bf h}_{_U}^\dagger{\bf h}_{_U}
+s_{_\beta}^2{\bf h}_{_D}^\dagger{\bf h}_{_D}){\cal Z}_{_Q}\Big)^{IJ}
\Big({\cal Z}_{_Q}^\dagger{\bf h}_{_U}^\dagger\Big)^{J3}\Big]
Q_9(x_{_{Q_I}},x_{_\mu},x_{_{Q_J}})
\nonumber\\
&&\hspace{1.3cm}
-{c_{_\beta}^2\over64\pi^2x_{_{H}}}\Big[{\bf h}_{_U}\Big(s_{_\beta}^2
{\bf h}_{_U}^\dagger{\bf h}_{_U}+c_{_\beta}^2{\bf h}_{_D}^\dagger{\bf h}_{_D}
\Big){\bf h}_{_U}^\dagger\Big]_{33}
+{1\over64\pi^2}\sum\limits_{q=U,D}{\cal X}_{_{q,IJK}}^{^S}
U_1(x_{_\mu},x_{_{Q_I}},x_{_{q_J}},x_{_{Q_K}})
\nonumber\\
&&\hspace{1.3cm}
+{1\over72\pi^2}{\cal X}_{_{U,IJK}}^{^R}
\Big[g_1^2U_1(x_{_i},x_{_{U_I}},x_{_{Q_J}},x_{_{U_K}})
+3g_3^2U_1(x_{_3},x_{_{U_I}},x_{_{Q_J}},x_{_{U_K}})\Big]
\nonumber\\
&&\hspace{1.3cm}
+{1\over144\pi^2}g_1^4{\cal Z}_{_U}^{\dagger3I}{\cal Z}_{_U}^{I3}
\Big[c_{_\beta}^2{\cal C}_{_{122}}^2(x_{_\mu},x_{_1},x_{_{U_I}})
+s_{_\beta}^2{\cal C}_{_{122}}^1(x_{_\mu},x_{_1},x_{_{U_I}})
+2\xi_{_{H1}}^S{\cal C}_{_{122}}^1(x_{_\mu},x_{_1},x_{_{U_I}})\Big]
\nonumber\\
&&\hspace{1.3cm}
+\sum\limits_{i=1,2}{F_ig_i^2\over128\pi^2}
\Big(h_{_U}{\cal Z}_{_Q}\Big)_{3I}
\Big({\cal Z}_{_Q}^\dagger h_{_U}^\dagger\Big)_{I3}
\Big[c_{_\beta}^2x_{_\mu}{\cal C}_{_{212}}^1(x_{_\mu},
x_{_i},x_{_{Q_I}})+s_{_\beta}^2{\cal C}_{_{212}}^2(x_{_\mu},x_{_i},
x_{_{Q_I}})\nonumber\\
&&\hspace{1.3cm}+2\xi_{_{Hi}}^S{\cal C}_{_{212}}^1(x_{_\mu},x_{_i},x_{_{Q_I}})\Big]
-{1\over48\pi^2}g_1^2s_{_\beta}\Gamma_{_{U,IJ}}^{^T}U_5(x_{_\mu},x_{_1},x_{_{Q_I}},
x_{_{U_J}})
\;,\nonumber\\
&&C_{_{t\Phi3}}=i{c_{_\beta}^2\over72\pi^2}{\bf Im}\Big[{\cal Z}_{_U}^{\dagger3I}
\Big({\cal Z}_{_U}{\bf h}_{_U}{\bf h}_{_U}^\dagger{\cal Z}_{_U}^\dagger
\Big)^{IJ}{\cal Z}_{_U}^{J3}\Big]\Big[g_1^2{\cal C}_{_{211}}^1
(x_{_{U_I}},x_{_1},x_{_{U_J}})
+3g_3^2{\cal C}_{_{211}}^1(x_{_{U_I}},x_{_3},x_{_{U_J}})\Big]
\nonumber\\
&&\hspace{1.3cm}
+{i\over16\pi^2}{\bf Im}\Big[\Big({\bf h}_{_U}{\cal Z}_{_Q}\Big)^{3I}
\Big({\cal Z}_{_Q}^\dagger(c_{_\beta}^2{\bf h}_{_U}^\dagger{\bf h}_{_U}
+s_{_\beta}^2{\bf h}_{_D}^\dagger{\bf h}_{_D}){\cal Z}_{_Q}\Big)^{IJ}
\Big({\cal Z}_{_Q}^\dagger{\bf h}_{_U}^\dagger\Big)^{J3}\Big]
{\cal C}_{_{211}}^1(x_{_{Q_I}},x_{_\mu},x_{_{Q_J}})
\nonumber\\
&&\hspace{1.3cm}
+i{1\over36\pi^2}
{\cal X}_{_{U,IJK}}^{^A}\Big[g_1^2{\cal D}_{_{1211}}^1(x_{_1},
x_{_{U_I}},x_{_{Q_J}},x_{_{U_K}})
+3g_3^2{\cal D}_{_{1211}}^1(x_{_3},
x_{_{U_I}},x_{_{Q_J}},x_{_{U_K}})\Big]
\nonumber\\
&&\hspace{1.3cm}
+i{1\over32\pi^2}\sum\limits_{q=U,D}{\cal X}_{_{q,IJK}}^{^B}
{\cal D}_{_{1211}}^1(x_{_\mu},x_{_{Q_I}},x_{_{q_J}},x_{_{Q_K}})
+{i\over36\pi^2}g_1^4\xi_{_{H1}}^B{\cal Z}_{_U}^{\dagger3I}{\cal Z}_{_U}^{I3}
{\cal C}_{_{121}}^0(x_{_\mu},x_{_1},x_{_{U_I}})
\nonumber\\
&&\hspace{1.3cm}
+i\sum\limits_{i=1,2}{F_ig_i^2\over32\pi^2}\xi_{_{Hi}}^B
\Big(h_{_U}{\cal Z}_{_Q}\Big)_{3I}\Big({\cal Z}_{_Q}^\dagger h_{_U}^\dagger\Big)_{I3}
{\cal C}_{_{211}}^0(x_{_\mu},x_{_i},x_{_{Q_I}})
\nonumber\\
&&\hspace{1.3cm}
+{i\over48\pi^2}g_1^2s_{_\beta}\Big[2\Gamma_{_{U,IJ}}^{^D}
{\cal D}_{_{1111}}^0(x_{_\mu},x_{_1},x_{_{Q_I}},
x_{_{U_J}})-\Gamma_{_{U,IJ}}^{^C}U_5(x_{_\mu},x_{_{U_J}},x_{_{Q_I}},
x_{_1})\Big]
\;.\nonumber\\
\label{wcoe3}
\end{eqnarray}
Just as in the appendix \ref{aop1}, the sum with the generation
indices $I,\;J,\;K$ is implied.


\section{The Wilson coefficients for operators ${\cal O}_{_{q\Phi i}}\;
(i=1,\;\cdots\;6)$\label{aop3}}

The Wilson coefficients for those operators
are written as
\begin{eqnarray}
&&C_{_{q\Phi1}}={1\over32\pi^2x_{_H}}s_{_\beta}^2c_{_\beta}^2
\Big[\Big({\bf h}_{_U}^{\dagger}{\bf h}_{_U}{\bf h}_{_U}^{\dagger}
{\bf h}_{_U}\Big)_{33}-\Big({\bf h}_{_D}^{\dagger}{\bf h}_{_D}
{\bf h}_{_D}^{\dagger}{\bf h}_{_D}\Big)_{33}\Big]
\Big(1-\ln x_{_H}\Big)
\nonumber\\
&&\hspace{1.3cm}
+\sum\limits_{q=U,D}(-1)^{{1\over2}-T_Z^q}{\cal X}_{_{q,IJK}}^{^T}\Big[
\sum\limits_{i=1,2}{F_i^2g_i^2\over2304\pi^2}
{\cal D}_{_{1121}}^1(x_{_i},x_{_{Q_I}},x_{_{q_J}},x_{_{Q_K}})
\nonumber\\
&&\hspace{1.3cm}
+{g_3^2\over48\pi^2}
{\cal D}_{_{1121}}^1(x_{_3},x_{_{Q_I}},x_{_{q_J}},x_{_{Q_K}})\Big]
-{1\over64\pi^2}\sum\limits_{q=U,D}(-1)^{{1\over2}-T_Z^q}{\cal X}_{_{q,IJK}}^{^U}
{\cal D}_{_{1121}}^1(x_{_\mu},x_{_{q_I}},x_{_{Q_J}},x_{_{q_K}})
\nonumber\\
&&\hspace{1.3cm}
+{g_1^4\over2304\pi^2}(c_{_\beta}^2-s_{_\beta}^2){\cal Z}_{_Q}^{3I}
{\cal Z}_{_Q}^{\dagger I3}Q_8(x_{_\mu},x_{_1},x_{_{Q_I}})
\nonumber\\
&&\hspace{1.3cm}
+\sum\limits_{i=1,2}{F_ig_i^2\over256\pi^2}
\Big({\bf h}_{_U}^\dagger{\cal Z}_{_U}^\dagger\Big)_{3I}
\Big({\cal Z}_{_U}{\bf h}_{_U}\Big)_{I3}\Big[s_{_\beta}^2Q_{10}(x_{_\mu},
x_{_i},x_{_{U_I}})-4\xi_{_{Hi}}^S
{\cal C}_{_{211}}^0(x_{_\mu},x_{_i},x_{_{U_I}})
\nonumber\\
&&\hspace{1.3cm}
+\eta_{_{Hi}}^RQ_7(x_{_i},x_{_\mu},x_{_{U_I}})\Big]
-\sum\limits_{i=1,2}{F_ig_i^2\over256\pi^2}
\Big({\bf h}_{_D}^\dagger{\cal Z}_{_D}^\dagger\Big)_{3I}
\Big({\cal Z}_{_D}{\bf h}_{_D}\Big)_{I3}\Big[c_{_\beta}^2
Q_{10}(x_{_\mu},x_{_i},x_{_{D_I}})
\nonumber\\
&&\hspace{1.3cm}
-4\xi_{_{Hi}}^S{\cal C}_{_{211}}^0(x_{_\mu},x_{_i},x_{_{D_I}})
+\eta_{_{Hi}}^SQ_7(x_{_i},x_{_\mu},
x_{_{D_I}})\Big]
\nonumber\\
&&\hspace{1.3cm}
+{g_1^2\over256\pi^2}\sum\limits_{q=U,D}Y_{_q}^B\Big[2\Gamma_{_{q,IJ}}^{^{W}}
{\cal D}_{_{1111}}^0(x_{_\mu},x_{_1},x_{_{q_I}},x_{_{Q_J}})
-\Gamma_{_{q,IJ}}^{^{V,1}}U_5(x_{_\mu},x_{_{q_I}},x_{_1}
,x_{_{Q_J}})\Big]
\nonumber\\
&&\hspace{1.3cm}
-{3g_2^2\over128\pi^2}\sum\limits_{q=U,D}\Big[2\Gamma_{_{q,IJ}}^{^{W}}
{\cal D}_{_{1111}}^0(x_{_\mu},x_{_2},x_{_{q_I}},x_{_{Q_J}})
-\Gamma_{_{q,IJ}}^{^{V,2}}U_5(x_{_\mu},x_{_{q_I}}^2,x_{_2}
,x_{_{Q_J}})\Big]
\;,\nonumber\\
&&C_{_{q\Phi2}}=\sum\limits_{q=U,D}
{\cal X}_{_{q,IJK}}^{^T}\Big[\sum\limits_{i=1,2}(-1)^i{F_i^2g_i^2\over2304\pi^2}
{\cal D}_{_{1121}}^1(x_{_i},x_{_{Q_I}},x_{_{q_J}},x_{_{Q_K}})
\nonumber\\
&&\hspace{1.3cm}
-{g_3^2\over48\pi^2}
{\cal D}_{_{1121}}^1(x_{_3},x_{_{Q_I}},x_{_{q_J}},x_{_{Q_K}})\Big]
+{g_2^4\over128\pi^2}(s_{_\beta}^2-c_{_\beta}^2){\cal Z}_{_Q}^{3I}
{\cal Z}_{_Q}^{\dagger I3}Q_8(x_{_\mu},x_{_2},x_{_{Q_I}})
\nonumber\\
&&\hspace{1.3cm}
-{g_1^2g_2^2\over384\pi^2}(s_{_\beta}^2-c_{_\beta}^2){\cal Z}_{_Q}^{3I}
{\cal Z}_{_Q}^{\dagger I3}\Big[U_2(x_{_\mu},x_{_1},x_{_2},x_{_{Q_I}})
+\sqrt{x_{_1}x_{_2}}\cos(\varphi_2-\varphi_1)U_3(x_{_\mu},x_{_1},x_{_2},x_{_{Q_I}})\Big]
\nonumber\\
&&\hspace{1.3cm}
+\sum\limits_{i=1,2}(-1)^i{g_i^2\over256\pi^2}\Big({\bf h}_{_U}^\dagger
{\cal Z}_{_U}^\dagger\Big)_{3I}
\Big({\cal Z}_{_U}{\bf h}_{_U}\Big)_{I3}\Big[s_{_\beta}^2Q_{10}(x_{_\mu},
x_{_i},x_{_{U_I}})
-4\xi_{_{Hi}}^S{\cal C}_{_{211}}^0(x_{_\mu},x_{_i},x_{_{U_I}})
\nonumber\\
&&\hspace{1.3cm}+\eta_{_{Hi}}^RQ_7(x_{_i},x_{_\mu},x_{_{U_I}})\Big]
+\sum\limits_{i=1,2}(-1)^i{g_i^2\over256\pi^2}
\Big({\bf h}_{_D}^\dagger{\cal Z}_{_D}^\dagger\Big)_{3I}
\Big({\cal Z}_{_D}{\bf h}_{_D}\Big)_{I3}\Big[c_{_\beta}^2Q_{10}(x_{_\mu},
x_{_i},x_{_{D_I}})\nonumber\\
&&\hspace{1.3cm}-4\xi_{_{Hi}}^S
{\cal C}_{_{211}}^0(x_{_\mu},x_{_i},x_{_{D_I}})
+\eta_{_{Hi}}^SQ_7(x_{_i},x_{_\mu},
x_{_{D_I}})\Big]
\nonumber\\
&&\hspace{1.3cm}
-{g_1^2\over256\pi^2}\sum\limits_{q=U,D}(-1)^{{1\over2}-T_Z^q}Y_{_q}^B
\Big[2\Gamma_{_{q,IJ}}^{^{W}}
{\cal D}_{_{1111}}^0(x_{_\mu},x_{_1},x_{_{q_I}},x_{_{Q_J}})
-\Gamma_{_{q,IJ}}^{^{V,1}}U_5(x_{_\mu},x_{_{q_I}},x_{_1}
,x_{_{Q_J}})\Big]
\nonumber\\
&&\hspace{1.3cm}
-{g_2^2\over128\pi^2}\sum\limits_{q=U,D}(-1)^{{1\over2}-T_Z^q}\Big[2\Gamma_{_{q,IJ}}^{^{W}}
{\cal D}_{_{1111}}^0(x_{_\mu},x_{_2},x_{_{q_I}},x_{_{Q_J}})
-\Gamma_{_{q,IJ}}^{^{V,2}}U_5(x_{_\mu},x_{_{q_I}},x_{_2},x_{_{Q_J}})\Big]
\;,\nonumber\\
&&C_{_{q\Phi3}}=3{g_1^2+g_2^2\over128\pi^2}s_{_\beta}^2c_{_\beta}^2
(s_{_\beta}^2-c_{_\beta}^2)\Big[\Big({\bf h}_{_U}^\dagger{\bf h}_{_U}\Big)_{33}
-\Big({\bf h}_{_D}^\dagger{\bf h}_{_D}\Big)_{33}\Big]{1+\ln x_{_H}\over x_{_H}}
\nonumber\\
&&\hspace{1.3cm}+{1\over256\pi^2x_{_H}}\Big(g_1^2-6s_{_\beta}^2c_{_\beta}^2(g_1^2
+g_2^2)\Big)\Big[c_{_\beta}^2\Big({\bf h}_{_U}^\dagger{\bf h}_{_U}\Big)_{33}
+s_{_\beta}^2\Big({\bf h}_{_D}^\dagger{\bf h}_{_D}\Big)_{33}\Big]
\nonumber\\
&&\hspace{1.3cm}
+{g_1^2(s_{_\beta}^2-c_{_\beta}^2)\over6912\pi^2}
{\cal Z}_{_Q}^{3I}{\cal Z}_{_Q}^{\dagger I3}
\Big[\sum\limits_{i=1,2}F_i^3g_i^2{\cal B}_{_{3,1}}^1(
x_{_{Q_I}},x_{_1})
+48g_3^2{\cal B}_{_{3,1}}^1(x_{_{Q_I}},x_{_3})\Big]
\nonumber\\
&&\hspace{1.3cm}+{1\over2304\pi^2}\Lambda_{_{IJ}}^S
\Big[\sum\limits_{i=1,2}F_i^2g_i^2Q_9(x_{_{Q_I}},x_{_i},x_{_{Q_J}})
+48g_3^2Q_9(x_{_{Q_I}},x_{_3},x_{_{Q_J}})\Big]
\nonumber\\
&&\hspace{1.3cm}
+{1\over128\pi^2}\sum\limits_{q=U,D}\Big[Y_{_q}^Bg_1^2(s_{_\beta}^2-c_{_\beta}^2)
\Big({\bf h}_{_q}^{\dagger}{\cal Z}_{_q}^{\dagger}\Big)_{3I}
\Big({\cal Z}_{_q}{\bf h}_{_q}\Big)_{I3}{\cal B}_{_{3,1}}^1(
x_{_{q_I}},x_{_\mu})+2\Lambda_{_{q,IJ}}^UQ_9(x_{_{q_I}},x_{_\mu},x_{_{q_J}})\Big]
\nonumber\\
&&\hspace{1.3cm}
-{1\over64\pi^2x_{_H}}s_{_\beta}^2c_{_\beta}^2
\Big[\Big({\bf h}_{_U}^{\dagger}{\bf h}_{_U}{\bf h}_{_U}^{\dagger}
{\bf h}_{_U}\Big)_{33}+\Big({\bf h}_{_U}^{\dagger}{\bf h}_{_U}
{\bf h}_{_U}^{\dagger}{\bf h}_{_U}\Big)_{33}\Big]
\nonumber\\
&&\hspace{1.3cm}
+\sum\limits_{q=U,D}{\cal X}_{_{q,IJK}}^{^T}\Big[\sum\limits_{i=1,2}
{F_i^3g_i^2\over2304\pi^2}U_1(x_{_i},x_{_{Q_I}},x_{_{q_J}},x_{_{Q_K}})
+{g_3^2\over48\pi^2}U_1(x_{_3},x_{_{Q_I}},x_{_{U_J}},x_{_{Q_K}})\Big]
\nonumber\\
&&\hspace{1.3cm}
+{g_1^4\over2304\pi^2}{\cal Z}_{_Q}^{3I}{\cal Z}_{_Q}^{\dagger I3}
\Big[{\cal C}_{_{122}}^2(x_{_\mu},
x_{_1},x_{_{Q_I}})+\eta_{_{H1}}^T{\cal C}_{_{122}}^1(x_{_\mu},
x_{_1},x_{_{Q_I}})\Big]
\nonumber\\
&&\hspace{1.3cm}
+\sum\limits_{i=1,2}{F_ig_i^2\over256\pi^2}
\Big({\bf h}_{_U}^\dagger{\cal Z}_{_U}^\dagger\Big)_{3I}
\Big({\cal Z}_{_U}{\bf h}_{_U}\Big)_{I3}
\Big[\eta_{_{Hi}}^R{\cal C}_{_{212}}^1(x_{_\mu},
x_{_i},x_{_{U_I}})
+s_{_\beta}^2{\cal C}_{_{212}}^2(x_{_\mu}
,x_{_i},x_{_{U_I}})\Big]
\nonumber\\
&&\hspace{1.3cm}
+\sum\limits_{i=1,2}{F_ig_i^2\over256\pi^2}
\Big({\bf h}_{_D}^\dagger{\cal Z}_{_D}^\dagger\Big)_{3I}
\Big({\cal Z}_{_D}{\bf h}_{_D}\Big)_{I3}
\Big[\eta_{_{Hi}}^S{\cal C}_{_{212}}^1(x_{_\mu},
x_{_i},x_{_{D_I}})
+c_{_\beta}^2{\cal C}_{_{212}}^2(x_{_\mu}
,x_{_i},x_{_{D_I}})\Big]
\nonumber\\
&&\hspace{1.3cm}
+{1\over64\pi^2}\sum\limits_{q=U,D}{\cal X}_{_{q,IJK}}^{^U}
U_1(x_{_\mu},x_{_{q_I}},x_{_{Q_J}},x_{_{q_K}})
+{g_1^2\over256\pi^2}\sum\limits_{q=U,D}(-1)^{{1\over2}-T_Z^q}Y_{_q}^B
\Gamma_{_{q,IJ}}^{^{V,1}}U_5(x_{_\mu},x_{_1},x_{_{q_I}},x_{_{Q_J}})
\nonumber\\
&&\hspace{1.3cm}
-{3g_2^2\over128\pi^2}\sum\limits_{q=U,D}(-1)^{{1\over2}-T_Z^q}
\Gamma_{_{q,IJ}}^{^{V,2}}U_5(x_{_\mu},x_{_2},x_{_{q_I}},x_{_{Q_J}})
\;,\nonumber\\
&&C_{_{q\Phi4}}=-{g_1^2+g_2^2\over128\pi^2}s_{_\beta}^2c_{_\beta}^2
(s_{_\beta}^2-c_{_\beta}^2)\Big[\Big({\bf h}_{_U}^\dagger{\bf h}_{_U}\Big)_{33}
+\Big({\bf h}_{_D}^\dagger{\bf h}_{_D}\Big)_{33}\Big]{1+\ln x_{_H}\over x_{_H}}
\nonumber\\
&&\hspace{1.3cm}-{1\over256\pi^2x_{_H}}\Big(g_2^2-2s_{_\beta}^2c_{_\beta}^2(g_1^2
+g_2^2)\Big)\Big[c_{_\beta}^2\Big({\bf h}_{_U}^\dagger{\bf h}_{_U}\Big)_{33}
-s_{_\beta}^2\Big({\bf h}_{_D}^\dagger{\bf h}_{_D}\Big)_{33}\Big]
\nonumber\\
&&\hspace{1.3cm}
-{g_2^2\over2304\pi^2}(s_{_\beta}^2-c_{_\beta}^2)
{\cal Z}_{_Q}^{3I}{\cal Z}_{_Q}^{\dagger I3}\Big[
\sum\limits_{i=1,2}(-1)^iF_i^2g_i^2{\cal B}_{_{3,1}}^1(
x_{_{Q_I}},x_{_i})
-48g_3^2{\cal B}_{_{3,1}}^1(x_{_{Q_I}},x_{_3})\Big]
\nonumber\\
&&\hspace{1.3cm}+{1\over2304\pi^2}\Lambda_{_{IJ}}^S
\Big[\sum\limits_{i=1,2}(-1)^{i-1}F_i^2g_i^2Q_9(x_{_{Q_I}},x_{_i},x_{_{Q_J}})
+48g_3^2Q_9(x_{_{Q_I}},x_{_3},x_{_{Q_J}})\Big]
\nonumber\\
&&\hspace{1.3cm}
+{1\over2304\pi^2}\sum\limits_{q=U,D}(-1)^{{1\over2}-T_Z^q}
{\cal X}_{_{q,IJK}}^{^T}\Big[\sum\limits_{i=1,2}
(-1)^{i}F_i^2g_i^2U_1(x_{_i},x_{_{Q_I}},x_{_{q_J}},x_{_{Q_K}})
\nonumber\\
&&\hspace{1.3cm}
-48g_3^2U_1(x_{_3},x_{_{Q_I}},x_{_{q_J}},x_{_{Q_K}})\Big]
-{g_2^2\over384\pi^2}{\cal Z}_{_Q}^{3I}{\cal Z}_{_Q}^{\dagger I3}
\bigg\{3g_2^2\Big[{\cal C}_{_{122}}^2(x_{_\mu},
x_{_2},x_{_{Q_I}})+\eta_{_{H2}}^T{\cal C}_{_{122}}^1(x_{_\mu},
x_{_2},x_{_{Q_I}})\Big]
\nonumber\\
&&\hspace{1.3cm}
-g_1^2\Big[{\cal D}_{_{1112}}^2(x_{_\mu},x_{_1},x_{_2},x_{_{Q_I}})
+\Big(\sqrt{x_{_1}x_{_2}}\cos(\varphi_2-\varphi_1)
+2[\xi_{_{H1}}^S+\xi_{_{H2}}^S]\Big)
{\cal D}_{_{1112}}^1(x_{_\mu},x_{_1},x_{_2},x_{_{Q_I}})\Big]\bigg\}
\nonumber\\
&&\hspace{1.3cm}
+\sum\limits_{i=1,2}(-1)^i{g_i^2\over256\pi^2}
\Big({\bf h}_{_U}^\dagger{\cal Z}_{_U}^\dagger\Big)_{3I}
\Big({\cal Z}_{_U}{\bf h}_{_U}\Big)_{I3}
\Big[\eta_{_{Hi}}^R{\cal C}_{_{212}}^1(x_{_\mu},
x_{_i},x_{_{U_I}})+s_{_\beta}^2{\cal C}_{_{212}}^2(x_{_\mu}
,x_{_i},x_{_{U_I}})\Big]
\nonumber\\
&&\hspace{1.3cm}
-\sum\limits_{i=1,2}(-1)^i{g_1^2\over256\pi^2}
\Big({\bf h}_{_D}^\dagger{\cal Z}_{_D}^\dagger\Big)_{3I}
\Big({\cal Z}_{_D}{\bf h}_{_D}\Big)_{I3}
\Big[\eta_{_{Hi}}^S{\cal C}_{_{212}}^1(x_{_\mu},
x_{_i},x_{_{D_I}})\nonumber\\
&&\hspace{1.3cm}
+c_{_\beta}^2{\cal C}_{_{212}}^2(x_{_\mu}
,x_{_i},x_{_{D_I}})\Big]
-{g_1^2\over256\pi^2}\sum\limits_{q=U,D}Y_{_q}^B\Gamma_{_{q,IJ}}^{^{V,1}}
U_5(x_{_\mu},x_{_1},x_{_{q_I}},x_{_{Q_J}})
\nonumber\\
&&\hspace{1.3cm}
-{g_2^2\over128\pi^2}\sum\limits_{q=U,D}\Gamma_{_{q,IJ}}^{^{V,2}}
U_5(x_{_\mu},x_{_2},x_{_{q_I}},x_{_{Q_J}})
\;,\nonumber\\
&&C_{_{q\Phi5}}={1\over2304\pi^2}\Big[\Lambda_{_{IJ}}^B\Big(
\sum\limits_{i=1,2}g_i^2F_i^3{\cal C}_{_{112}}^1(x_{_{Q_I}},x_{_i},x_{_{Q_J}})
+48g_3^2{\cal C}_{_{112}}^1(x_{_{Q_I}},x_{_3},x_{_{Q_J}})\Big)
+36\sum\limits_{q=U,D}\Lambda_{_{q,IJ}}^D{\cal C}_{_{112}}^1(x_{_{q_I}}
,x_{_\mu},x_{_{q_J}})\Big]
\nonumber\\
&&\hspace{1.3cm}
+{i\over1152\pi^2}\sum\limits_{q=U,D}\sum\limits_{i=1,2}
{\cal X}_{_{q,IJK}}^{^C}\Big[F_i^3g_i^2
{\cal D}_{_{1211}}^1(x_{_i},x_{_{Q_I}},x_{_{q_J}},x_{_{Q_K}})
\nonumber\\
&&\hspace{1.3cm}
+48g_3^2{\cal D}_{_{1211}}^1(x_{_3},x_{_{Q_I}},x_{_{q_J}},x_{_{Q_K}})\Big]
+{i\over32\pi^2}\sum\limits_{q=U,D}{\cal X}_{_{q,IJK}}^{^D}
{\cal D}_{_{1211}}^1(x_{_\mu},x_{_{q_I}},x_{_{Q_J}},x_{_{q_K}})
\nonumber\\
&&\hspace{1.3cm}
-i{g_1^4\over288\pi^2}\xi_{_{H1}}^B
{\cal Z}_{_Q}^{3I}{\cal Z}_{_Q}^{\dagger I3}
{\cal C}_{_{121}}^0(x_{_\mu},x_{_1},x_{_{Q_I}})
\nonumber\\
&&\hspace{1.3cm}
-{i\over64\pi^2}\sum\limits_{q=U,D}\sum\limits_{i=1,2}F_ig_i^2\xi_{_{Hi}}^B
\Big({\bf h}_{_q}^\dagger{\cal Z}_{_q}^\dagger\Big)_{3I}
\Big({\cal Z}_{_q}{\bf h}_{_q}\Big)_{I3}
{\cal C}_{_{211}}^0(x_{_\mu},x_{_i},x_{_{q_I}})
\nonumber\\
&&\hspace{1.3cm}
+i{g_1^2\over256\pi^2}\sum\limits_{q=U,D}(-1)^{{1\over2}-T_Z^q}Y_{_q}^B
\Big[\Gamma_{_{q,IJ}}^{^{E,1}}U_5(x_{_\mu},x_{_{Q_J}},x_{_{q_I}},x_{_1})
-2\Gamma_{_{q,IJ}}^{^{F}}{\cal D}_{_{1111}}^0(x_{_\mu},x_{_1},x_{_{q_I}},x_{_{Q_J}})\Big]
\nonumber\\
&&\hspace{1.3cm}
-i{3g_2^2\over128\pi^2}\sum\limits_{q=U,D}(-1)^{{1\over2}-T_Z^q}
\Big[\Gamma_{_{q,IJ}}^{^{E,2}}U_5(x_{_\mu},x_{_{Q_J}},x_{_{q_I}},x_{_2})
-2\Gamma_{_{q,IJ}}^{^{F}}{\cal D}_{_{1111}}^0(x_{_\mu},x_{_2},x_{_{q_I}},x_{_{Q_J}})\Big]
\;,\nonumber\\
&&C_{_{q\Phi6}}={\Lambda_{_{IJ}}^B\over2304\pi^2}\Big(
\sum\limits_{i=1,2}(-1)^{i-1}g_i^2F_i^3{\cal C}_{_{112}}^1(x_{_{Q_I}},x_{_i},x_{_{Q_J}})
+48g_3^2{\cal C}_{_{112}}^1(x_{_{Q_I}},x_{_3},x_{_{Q_J}})\Big)
\nonumber\\
&&\hspace{1.3cm}
+{i\over1152\pi^2}\sum\limits_{q=U,D}(-1)^{{1\over2}-T_Z^q}
\Big[\sum\limits_{i=1,2}(-1)^iF_i^2g_i^2{\cal X}_{_{q,IJK}}^{^C}
{\cal D}_{_{1211}}^1(x_{_1},x_{_{Q_I}},x_{_{q_J}},x_{_{Q_K}})
\nonumber\\
&&\hspace{1.3cm}
-48g_3^2{\cal X}_{_{q,IJK}}^{^C}
{\cal D}_{_{1211}}^1(x_{_3},x_{_{Q_I}},x_{_{q_J}},x_{_{Q_K}})\Big]
+i{g_2^4\over16\pi^2}\xi_{_{H2}}^B
{\cal Z}_{_Q}^{3I}{\cal Z}_{_Q}^{\dagger I3}
{\cal C}_{_{121}}^0(x_{_\mu},x_{_2},x_{_{Q_I}})
\nonumber\\
&&\hspace{1.3cm}
-i{g_1^2g_2^2\over384\pi^2}{\cal Z}_{_Q}^{3I}{\cal Z}_{_Q}^{\dagger I3}
\Big[4\Big(\xi_{_{H1}}^B+\xi_{_{H2}}^B\Big)
{\cal D}_{_{1111}}^0(x_{_\mu},x_{_1},x_{_2},x_{_{Q_I}})
\nonumber\\
&&\hspace{1.3cm}
-\Big((c_{_\beta}^2-s_{_\beta}^2)\sqrt{x_{_1}x_{_2}}\sin(\varphi_2-\varphi_1)
-2[\xi_{_{H1}}^B-\xi_{_{H2}}^B]\Big)U_4(x_{_\mu},x_{_1},x_{_2},x_{_{Q_I}})\Big]
\nonumber\\
&&\hspace{1.3cm}
-{1\over64\pi^2}\sum\limits_{q=U,D}(-1)^{{1\over2}-T_Z^q}
\sum\limits_{i=1,2}(-1)^i
g_i^2\xi_{_{Hi}}^B
\Big({\bf h}_{_U}^\dagger{\cal Z}_{_U}^\dagger\Big)_{3I}
\Big({\cal Z}_{_U}{\bf h}_{_U}\Big)_{I3}
{\cal C}_{_{211}}^0(x_{_\mu},x_{_i},x_{_{U_I}})
\nonumber\\
&&\hspace{1.3cm}
-i{g_1^2\over256\pi^2}\sum\limits_{q=U,D}Y_{_q}^B\Big[\Gamma_{_{q,IJ}}^{^{E,1}}
U_5(x_{_\mu},x_{_{Q_J}},x_{_{q_I}},x_{_1})
-2\Gamma_{_{q,IJ}}^{^{F}}{\cal D}_{_{1111}}^0(x_{_\mu},x_{_1},x_{_{q_I}},x_{_{Q_J}})\Big]
\nonumber\\
&&\hspace{1.3cm}
-i{g_2^2\over128\pi^2}\sum\limits_{q=U,D}\Big[\Gamma_{_{q,IJ}}^{^{E,2}}
U_5(x_{_\mu},x_{_{Q_J}},x_{_{q_I}},x_{_2})
-2\Gamma_{_{q,IJ}}^{^{F}}{\cal D}_{_{1111}}^0(x_{_\mu},x_{_2}
,x_{_{q_I}},x_{_{Q_J}})\Big]
\;.\nonumber\\
\label{wcoe4}
\end{eqnarray}

\section{The coupling constants and loop functions \label{app4}}

\indent\indent
The loop functions are defined as
\begin{eqnarray}
&&P_{_H}(x,y)=-{\cal B}_{_{1,3}}^0(x,y)+3{\cal B}_{_{1,4}}^1(x,y)
-{2\over3}{\cal B}_{_{1,5}}^2(x,y)
\;,\nonumber\\
&&Q_1(x,y,z)={\cal C}_{_{122}}^1(x,y,z)+
y{\cal C}_{_{131}}^0(x,y,z)+z{\cal C}_{_{113}}^0(x,y,z)
\;,\nonumber\\
&&Q_2(x,y,z)={\cal C}_{_{122}}^2(x,y,z)+
y{\cal C}_{_{131}}^1(x,y,z)+z{\cal C}_{_{113}}^1(x,y,z)
\;,\nonumber\\
&&Q_3(x,y,z)={\cal C}_{_{122}}^1(x,y,z)+
x{\cal C}_{_{131}}^0(x,y,z)+z{\cal C}_{_{113}}^0(x,y,z)
\;,\nonumber\\
&&Q_4(x,y,z)={\cal C}_{_{122}}^2(x,y,z)+
x{\cal C}_{_{131}}^1(x,y,z)+z{\cal C}_{_{113}}^1(x,y,z)
\;,\nonumber\\
&&Q_5(x,y,z)={\cal C}_{_{122}}^2(x,y,z)+2{\cal C}_{_{131}}^2(x,y,z)
-4{\cal C}_{_{121}}^1(x,y,z)
\;,\nonumber\\
&&Q_6(x,y,z)={\cal C}_{_{122}}^1(x,y,z)+2{\cal C}_{_{131}}^1(x,y,z)
+2{\cal C}_{_{121}}^0(x,y,z)
\;,\nonumber\\
&&Q_7(x,y,z)={\cal C}_{_{122}}^1(x,y,z)+2{\cal C}_{_{131}}^1(x,y,z)
\;,\nonumber\\
&&Q_8(x,y,z)=4{\cal C}_{_{121}}^1(x,y,z)-2{\cal C}_{_{131}}^2(x,y,z)
-{\cal C}_{_{122}}^2(x,y,z)+y\Big(2{\cal C}_{_{131}}^1(x,y,z)
+{\cal C}_{_{122}}^1(x,y,z)\Big)\;\nonumber\\
&&Q_9(x,y,z)={\cal C}_{_{211}}^1(x,y,z)+{\cal C}_{_{112}}^1(x,y,z)
\;\nonumber\\
&&Q_{10}(x,y,z)={\cal C}_{_{212}}^2(x,y,z)+2{\cal C}_{_{311}}^2(x,y,z)
-4{\cal C}_{_{211}}^1(x,y,z)
\;\nonumber\\
&&U_1(x,y,z,w)=2{\cal D}_{_{1211}}^1(x,y,z,w)+{\cal D}_{_{1121}}^1(x,y,z,w)
\;,\nonumber\\
&&U_2(x,y,z,w)=4{\cal D}_{_{1111}}^1(x,y,z,w)
-{\cal D}_{_{1211}}^2(x,y,z,w)-{\cal D}_{_{1121}}^2(x,y,z,w)
-{\cal D}_{_{1112}}^2(x,y,z,w)\;\nonumber\\
&&U_3(x,y,z,w)={\cal D}_{_{1211}}^1(x,y,z,w)
+{\cal D}_{_{1121}}^1(x,y,z,w)+{\cal D}_{_{1112}}^1(x,y,z,w)
\;\nonumber\\
&&U_4(x,y,z,w)={\cal D}_{_{1211}}^1(x,y,z,w)
-{\cal D}_{_{1121}}^1(x,y,z,w)
\;\nonumber\\
&&U_5(x,y,z,w)={\cal D}_{_{1121}}^1(x,y,z,w)
+{\cal D}_{_{1112}}^1(x,y,z,w)
\label{fun}
\end{eqnarray}

The coupling constants are
\begin{eqnarray}
&&\xi_{_{Hi}}^R=\sqrt{x_{_\mu}x_{_i}}\Big((c_{_\beta}^2-s_{_\beta}^2)
{\bf Re}({\bf h}_{_U})_{33}\cos(\varphi_\mu+\varphi_i)
-{\bf Im}({\bf h}_{_U})_{33}\sin(\varphi_\mu+\varphi_i)\Big)
\;,\;(i=1,\;2)\;,\nonumber\\
&&\xi_{_{Hi}}^A=\sqrt{x_{_\mu}x_{_i}}\Big({\bf Im}({\bf h}_{_U})_{33}
(c_{_\beta}^2-s_{_\beta}^2)\cos(\varphi_\mu+\varphi_i)
+{\bf Re}({\bf h}_{_U})_{33}\sin(\varphi_\mu+\varphi_i)\Big)
\;,\;(i=1,\;2)\;,\nonumber\\
&&\xi_{_{Hi}}^S=s_{_\beta}c_{_\beta}\sqrt{x_{_\mu}x_{_i}}\cos(\varphi_\mu+\varphi_i)
\;,\;(i=1,\;2)\;,\nonumber\\
&&\xi_{_{Hi}}^B=s_{_\beta}c_{_\beta}\sqrt{x_{_\mu}x_{_i}}\sin(\varphi_\mu+\varphi_i)
\;,\;(i=1,\;2)\;,\nonumber\\
&&\eta_{_{Hi}}^R=c_{_\beta}^2x_{_\mu}+2s_{_\beta}c_{_\beta}\sqrt{x_{_\mu}x_{_i}}
\cos(\varphi_\mu+\varphi_i)
\;,\;(i=1,\;2)\;,\nonumber\\
&&\eta_{_{Hi}}^S=s_{_\beta}^2x_{_\mu}+2s_{_\beta}c_{_\beta}\sqrt{x_{_\mu}x_{_i}}
\cos(\varphi_\mu+\varphi_i)
\;,\;(i=1,\;2)\;,\nonumber\\
&&\eta_{_{Hi}}^T=x_{_i}+4s_{_\beta}c_{_\beta}\sqrt{x_{_\mu}x_{_i}}
\cos(\varphi_\mu+\varphi_i)
\;,\;(i=1,\;2)\;,\nonumber\\
&&\Lambda_{_{U,I}}^{^{R,i}}=\sqrt{x_{_\mu}x_{_i}}\bigg({\bf Re}\Big[{\cal Z}_{_Q}^{3I}
\Big({\cal Z}_{_Q}^\dagger{\bf h}_{_U}^\dagger\Big)^{I3}\Big]
\cos(\varphi_\mu+\varphi_i)
-{\bf Im}\Big[{\cal Z}_{_Q}^{3I}\Big({\cal Z}_{_Q}^\dagger
{\bf h}_{_U}^\dagger\Big)^{I3}\Big]\sin(\varphi_\mu+\varphi_i)\bigg)
\;,\;(i=1,\;{\rm or}\;2)\nonumber\\
&&\Lambda_{_{U,I}}^{^{A,i}}=\sqrt{x_{_\mu}x_{_i}}\bigg({\bf Re}\Big[{\cal Z}_{_Q}^{3I}
\Big({\cal Z}_{_Q}^\dagger
{\bf h}_{_U}^\dagger\Big)^{I3}\Big]\sin(\varphi_\mu+\varphi_i)
+{\bf Im}\Big[{\cal Z}_{_Q}^{3I}\Big({\cal Z}_{_Q}^\dagger
{\bf h}_{_U}^\dagger\Big)^{I3}\Big]\cos(\varphi_\mu+\varphi_i)
\bigg)\;,\;(i=1,\;{\rm or}\;2)\nonumber\\
&&\Lambda_{_{IJ}}^S={\bf Re}\Big[{\cal Z}_{_Q}^{3I}\Big(
{\cal Z}_{_Q}^\dagger(c_{_\beta}^2\Big({\bf h}_{_U}^\dagger{\bf h}_{_U}\Big)_{33}
+s_{_\beta}^2\Big({\bf h}_{_D}^\dagger{\bf h}_{_D})
{\cal Z}_{_Q}\Big)^{IJ}{\cal Z}_{_Q}^{\dagger J3}\Big]\;,
\nonumber\\
&&\Lambda_{_{IJ}}^B={\bf Im}\Big[{\cal Z}_{_Q}^{3I}\Big(
{\cal Z}_{_Q}^\dagger(c_{_\beta}^2\Big({\bf h}_{_U}^\dagger{\bf h}_{_U}\Big)_{33}
+s_{_\beta}^2\Big({\bf h}_{_D}^\dagger{\bf h}_{_D})
{\cal Z}_{_Q}\Big)^{IJ}{\cal Z}_{_Q}^{\dagger J3}\Big]\;,
\nonumber\\
&&\Lambda_{_{U,I}}^{^T}={\bf Re}\Big(({\bf h}_{_U}^\dagger
{\cal Z}_{_U}^\dagger)^{3I}{\cal Z}_{_U}^{I3}\Big)
\;,\;\;
\Lambda_{_{U,I}}^{^C}={\bf Im}\Big(({\bf h}_{_U}^\dagger
{\cal Z}_{_U}^\dagger)^{3I}{\cal Z}_{_U}^{I3}\Big)
\;,\nonumber\\
&&\Lambda_{_{Q,I}}^{^T}={\bf Re}\Big({\cal Z}_{_Q}^{3I}
({\cal Z}_{_Q}^\dagger{\bf h}_{_U}^\dagger)^{I3}\Big)
\;,\;\;\Lambda_{_{Q,I}}^{^C}={\bf Im}\Big({\cal Z}_{_Q}^{3I}
({\cal Z}_{_Q}^\dagger{\bf h}_{_U}^\dagger)^{I3}\Big)
\;,\nonumber\\
&&\Lambda_{_{U,IJ}}^U=c_{_\beta}^2{\bf Re}\Big[
\Big({\cal Z}_{_U}{\bf h}_{_U}\Big)^{\dagger 3I}\Big({\cal Z}_{_U}
{\bf h}_{_U}{\bf h}_{_U}^\dagger{\cal Z}_{_U}^\dagger\Big)^{IJ}
\Big({\cal Z}_{_U}{\bf h}_{_U}\Big)^{J3}\Big]\;,\nonumber\\
&&\Lambda_{_{U,IJ}}^D=c_{_\beta}^2{\bf Im}\Big[
\Big({\cal Z}_{_U}{\bf h}_{_U}\Big)^{\dagger 3I}\Big({\cal Z}_{_U}
{\bf h}_{_U}{\bf h}_{_U}^\dagger{\cal Z}_{_U}^\dagger\Big)^{IJ}
\Big({\cal Z}_{_U}{\bf h}_{_U}\Big)^{J3}\Big]\;,\nonumber\\
&&\Lambda_{_{D,IJ}}^U=s_{_\beta}^2{\bf Re}\Big[
\Big({\cal Z}_{_D}{\bf h}_{_D}\Big)^{\dagger 3I}\Big({\cal Z}_{_D}
{\bf h}_{_D}{\bf h}_{_D}^\dagger{\cal Z}_{_D}^\dagger\Big)^{IJ}
\Big({\cal Z}_{_D}{\bf h}_{_D}\Big)^{J3}\Big]\;,\nonumber\\
&&\Lambda_{_{D,IJ}}^D=s_{_\beta}^2{\bf Im}\Big[
\Big({\cal Z}_{_D}{\bf h}_{_D}\Big)^{\dagger 3I}\Big({\cal Z}_{_D}
{\bf h}_{_D}{\bf h}_{_D}^\dagger{\cal Z}_{_D}^\dagger\Big)^{IJ}
\Big({\cal Z}_{_D}{\bf h}_{_D}\Big)^{J3}\Big]\;,\nonumber\\
&&\Gamma_{_{U,IJ}}^{^{R,i}}=\sqrt{x_{_i}}\bigg({\bf Re}\Big[{\cal Z}_{_Q}^{3I}\Big({\cal Z}_{_U}
\hat{\bf A}_{_U}{\cal Z}_{_Q}\Big)^{\dagger IJ}{\cal Z}_{_U}^{J3}\Big]
\cos\varphi_i-{\bf Im}\Big[{\cal Z}_{_Q}^{3I}\Big({\cal Z}_{_U}
\hat{\bf A}_{_U}{\cal Z}_{_Q}\Big)^{\dagger IJ}{\cal Z}_{_U}^{J3}\Big]
\sin\varphi_i\bigg)\;,\;(i=1,\;{\rm or}\;3)\nonumber\\
&&\Gamma_{_{U,IJ}}^{^{A,i}}=\sqrt{x_{_i}}\bigg({\bf Re}\Big[{\cal Z}_{_Q}^{3I}\Big({\cal Z}_{_U}
\hat{\bf A}_{_U}{\cal Z}_{_Q}\Big)^{\dagger IJ}{\cal Z}_{_U}^{J3}\Big]
\sin\varphi_i+{\bf Im}\Big[{\cal Z}_{_Q}^{3I}\Big({\cal Z}_{_U}
\hat{\bf A}_{_U}{\cal Z}_{_Q}\Big)^{\dagger IJ}{\cal Z}_{_U}^{J3}\Big]
\cos\varphi_i\bigg)\;,\;(i=1,\;{\rm or}\;3)\nonumber\\
&&\Gamma_{_{U,IJ}}^{^S}=\sqrt{x_{_\mu}}\bigg({\bf Re}\Big[\Big({\bf h}_{_D}^\dagger{\cal Z}_{_D}
^\dagger\Big)^{3I}\Big({\cal Z}_{_D}
\hat{\bf A}_{_D}{\cal Z}_{_Q}\Big)^{IJ}\Big({\cal Z}_{_Q}^\dagger
{\bf h}_{_U}^\dagger\Big)^{J3}\Big]
\cos\varphi_\mu\nonumber\\
&&\hspace{1.3cm}-{\bf Im}\Big[\Big({\bf h}_{_D}^\dagger{\cal Z}_{_D}
^\dagger\Big)^{3I}\Big({\cal Z}_{_D}
\hat{\bf A}_{_D}{\cal Z}_{_Q}\Big)^{IJ}\Big({\cal Z}_{_Q}^\dagger
{\bf h}_{_U}^\dagger\Big)^{J3}\Big]
\sin\varphi_\mu\bigg)\;,\nonumber\\
&&\Gamma_{_{U,IJ}}^{^B}=\sqrt{x_{_\mu}}\bigg({\bf Re}\Big[\Big({\bf h}_{_D}^\dagger{\cal Z}_{_D}
^\dagger\Big)^{3I}\Big({\cal Z}_{_D}
\hat{\bf A}_{_D}{\cal Z}_{_Q}\Big)^{IJ}\Big({\cal Z}_{_Q}^\dagger
{\bf h}_{_U}^\dagger\Big)^{J3}\Big]
\sin\varphi_\mu\nonumber\\
&&\hspace{1.3cm}+{\bf Im}\Big[\Big({\bf h}_{_D}^\dagger{\cal Z}_{_D}
^\dagger\Big)^{3I}\Big({\cal Z}_{_D}
\hat{\bf A}_{_D}{\cal Z}_{_Q}\Big)^{IJ}\Big({\cal Z}_{_Q}^\dagger
{\bf h}_{_U}^\dagger\Big)^{J3}\Big]
\cos\varphi_\mu\bigg)\;,\nonumber\\
&&\Gamma_{_{U,IJ}}^{^T}={\bf Re}\Big[\Big({\bf h}_{_U}{\cal Z}_{_Q}\Big)_{3I}
\Big({\cal Z}_{_U}\hat{\bf A}_{_U}{\cal Z}_{_Q}\Big)^\dagger_{IJ}
{\cal Z}_{_U}^{J3}\Big]\Big[c_{_\beta}\sqrt{x_{_\mu}}\cos\varphi_\mu
+s_{_\beta}\sqrt{x_{_1}}\cos\varphi_1\Big]\nonumber\\
&&\hspace{1.5cm}
+{\bf Im}\Big[\Big({\bf h}_{_U}{\cal Z}_{_Q}\Big)_{3I}
\Big({\cal Z}_{_U}\hat{\bf A}_{_U}{\cal Z}_{_Q}\Big)^\dagger_{IJ}
{\cal Z}_{_U}^{J3}\Big]\Big[c_{_\beta}\sqrt{x_{_\mu}}\sin\varphi_\mu
-s_{_\beta}\sqrt{x_{_1}}\sin\varphi_1\Big],\;\nonumber\\
&&\Gamma_{_{U,IJ}}^{^U}=c_{_\beta}\sqrt{x_{_\mu}}\bigg\{{\bf Re}
\Big[\Big({\bf h}_{_U}{\cal Z}_{_Q}\Big)_{3I}
\Big({\cal Z}_{_U}\hat{\bf A}_{_U}{\cal Z}_{_Q}\Big)^\dagger_{IJ}
{\cal Z}_{_U}^{J3}\Big]\cos\varphi_\mu
+{\bf Im}\Big[\Big({\bf h}_{_U}{\cal Z}_{_Q}\Big)_{3I}
\Big({\cal Z}_{_U}\hat{\bf A}_{_U}{\cal Z}_{_Q}\Big)^\dagger_{IJ}
{\cal Z}_{_U}^{J3}\Big]\sin\varphi_\mu\bigg\}
,\;\nonumber\\
&&\Gamma_{_{U,IJ}}^{^C}={\bf Im}\Big[\Big({\bf h}_{_U}{\cal Z}_{_Q}\Big)_{3I}
\Big({\cal Z}_{_U}\hat{\bf A}_{_U}{\cal Z}_{_Q}\Big)^\dagger_{IJ}
{\cal Z}_{_U}^{J3}\Big]\Big[c_{_\beta}\sqrt{x_{_\mu}}\cos\varphi_\mu
+s_{_\beta}\sqrt{x_{_1}}\cos\varphi_1\Big]\nonumber\\
&&\hspace{1.5cm}-{\bf Re}\Big[\Big({\bf h}_{_U}{\cal Z}_{_Q}\Big)_{3I}
\Big({\cal Z}_{_U}\hat{\bf A}_{_U}{\cal Z}_{_Q}\Big)^\dagger_{IJ}
{\cal Z}_{_U}^{J3}\Big]\Big[c_{_\beta}\sqrt{x_{_\mu}}\sin\varphi_\mu
-s_{_\beta}\sqrt{x_{_1}}\sin\varphi_1\Big],\;\nonumber\\
&&\Gamma_{_{U,IJ}}^{^D}=c_{_\beta}\sqrt{x_{_\mu}}\bigg\{{\bf Im}
\Big[\Big({\bf h}_{_U}{\cal Z}_{_Q}\Big)_{3I}
\Big({\cal Z}_{_U}\hat{\bf A}_{_U}{\cal Z}_{_Q}\Big)^\dagger_{IJ}
{\cal Z}_{_U}^{J3}\Big]\cos\varphi_\mu
-{\bf Re}\Big[\Big({\bf h}_{_U}{\cal Z}_{_Q}\Big)_{3I}
\Big({\cal Z}_{_U}\hat{\bf A}_{_U}{\cal Z}_{_Q}\Big)^\dagger_{IJ}
{\cal Z}_{_U}^{J3}\Big]\sin\varphi_\mu\bigg\},\;\nonumber\\
&&\Gamma_{_{U,IJ}}^{^{V,i}}={\bf Re}\Big[\Big({\bf h}_{_U}^\dagger{\cal Z}_{_U}^\dagger\Big)_{3I}
\Big({\cal Z}_{_U}\hat{\bf A}_{_U}{\cal Z}_{_Q}\Big)_{IJ}{\cal Z}_{_Q}^{\dagger J3}\Big]
s_{_\beta}\Big(c_{_\beta}\sqrt{x_{_\mu}}\cos\varphi_\mu+s_{_\beta}\sqrt{x_{_i}}\cos\varphi_i\Big)
\nonumber\\&&\hspace{1.5cm}
-{\bf Im}\Big[\Big({\bf h}_{_U}^\dagger{\cal Z}_{_U}^\dagger\Big)_{3I}
\Big({\cal Z}_{_U}\hat{\bf A}_{_U}{\cal Z}_{_Q}\Big)_{IJ}{\cal Z}_{_Q}^{\dagger J3}\Big]
s_{_\beta}\Big(c_{_\beta}\sqrt{x_{_\mu}}\sin\varphi_\mu-s_{_\beta}\sqrt{x_{_i}}\sin\varphi_i\Big)
\;,(i=1,\;2)\;,\nonumber\\
&&\Gamma_{_{U,IJ}}^{^{E,i}}={\bf Re}\Big[\Big({\bf h}_{_U}^\dagger{\cal Z}_{_U}^\dagger\Big)_{3I}
\Big({\cal Z}_{_U}\hat{\bf A}_{_U}{\cal Z}_{_Q}\Big)_{IJ}{\cal Z}_{_Q}^{\dagger J3}\Big]
s_{_\beta}\Big(c_{_\beta}\sqrt{x_{_\mu}}\sin\varphi_\mu-s_{_\beta}\sqrt{x_{_i}}\sin\varphi_i\Big)
\nonumber\\&&\hspace{1.5cm}
+{\bf Im}\Big[\Big({\bf h}_{_U}^\dagger{\cal Z}_{_U}^\dagger\Big)_{3I}
\Big({\cal Z}_{_U}\hat{\bf A}_{_U}{\cal Z}_{_Q}\Big)_{IJ}{\cal Z}_{_Q}^{\dagger J3}\Big]
s_{_\beta}\Big(c_{_\beta}\sqrt{x_{_\mu}}\cos\varphi_\mu+s_{_\beta}\sqrt{x_{_i}}\cos\varphi_i\Big)
\;,(i=1,\;2)\;,\nonumber\\
&&\Gamma_{_{q,IJ}}^{^W}=s_{_\beta}c_{_\beta}\sqrt{x_{_\mu}}\bigg\{
{\bf Re}\Big[\Big({\bf h}_{_q}^\dagger{\cal Z}_{_q}^\dagger\Big)_{3I}
\Big({\cal Z}_{_q}\hat{\bf A}_{_q}{\cal Z}_{_Q}\Big)_{IJ}{\cal Z}_{_Q}^{\dagger J3}\Big]
\cos\varphi_\mu
\nonumber\\&&\hspace{1.5cm}
-{\bf Im}\Big[\Big({\bf h}_{_q}^\dagger{\cal Z}_{_q}^\dagger\Big)_{3I}
\Big({\cal Z}_{_q}\hat{\bf A}_{_q}{\cal Z}_{_Q}\Big)_{IJ}{\cal Z}_{_Q}^{\dagger J3}\Big]
\sin\varphi_\mu\bigg\}\;,\nonumber\\
&&\Gamma_{_{q,IJ}}^{^F}=s_{_\beta}c_{_\beta}\sqrt{x_{_\mu}}\bigg\{
{\bf Re}\Big[\Big({\bf h}_{_q}^\dagger{\cal Z}_{_q}^\dagger\Big)_{3I}
\Big({\cal Z}_{_q}\hat{\bf A}_{_q}{\cal Z}_{_Q}\Big)_{IJ}{\cal Z}_{_Q}^{\dagger J3}\Big]
\sin\varphi_\mu
\nonumber\\&&\hspace{1.5cm}
+{\bf Im}\Big[\Big({\bf h}_{_q}^\dagger{\cal Z}_{_q}^\dagger\Big)_{3I}
\Big({\cal Z}_{_q}\hat{\bf A}_{_U}{\cal Z}_{_Q}\Big)_{IJ}{\cal Z}_{_Q}^{\dagger J3}\Big]
\cos\varphi_\mu\bigg\}\;,\nonumber\\
&&\Gamma_{_{D,IJ}}^{^{V,i}}=-{\bf Re}\Big[\Big({\bf h}_{_D}^\dagger{\cal Z}_{_D}^\dagger\Big)_{3I}
\Big({\cal Z}_{_D}\hat{\bf A}_{_D}{\cal Z}_{_Q}\Big)_{IJ}{\cal Z}_{_Q}^{\dagger J3}\Big]
c_{_\beta}\Big(s_{_\beta}\sqrt{x_{_\mu}}\cos\varphi_\mu+c_{_\beta}\sqrt{x_{_i}}\cos\varphi_i\Big)
\nonumber\\&&\hspace{1.5cm}
+{\bf Im}\Big[\Big({\bf h}_{_D}^\dagger{\cal Z}_{_D}^\dagger\Big)_{3I}
\Big({\cal Z}_{_D}\hat{\bf A}_{_D}{\cal Z}_{_Q}\Big)_{IJ}{\cal Z}_{_Q}^{\dagger J3}\Big]
c_{_\beta}\Big(s_{_\beta}\sqrt{x_{_\mu}}\sin\varphi_\mu-c_{_\beta}\sqrt{x_{_i}}\sin\varphi_i\Big)\;,\nonumber\\
&&\Gamma_{_{D,IJ}}^{^{E,i}}=-{\bf Re}\Big[\Big({\bf h}_{_D}^\dagger{\cal Z}_{_D}^\dagger\Big)_{3I}
\Big({\cal Z}_{_D}\hat{\bf A}_{_D}{\cal Z}_{_Q}\Big)_{IJ}{\cal Z}_{_Q}^{\dagger J3}\Big]
c_{_\beta}\Big(s_{_\beta}\sqrt{x_{_\mu}}\sin\varphi_\mu-c_{_\beta}\sqrt{x_{_i}}\sin\varphi_i\Big)
\nonumber\\&&\hspace{1.5cm}
-{\bf Im}\Big[\Big({\bf h}_{_D}^\dagger{\cal Z}_{_D}^\dagger\Big)_{3I}
\Big({\cal Z}_{_D}\hat{\bf A}_{_D}{\cal Z}_{_Q}\Big)_{IJ}{\cal Z}_{_Q}^{\dagger J3}\Big]
c_{_\beta}\Big(s_{_\beta}\sqrt{x_{_\mu}}\cos\varphi_\mu+c_{_\beta}\sqrt{x_{_i}}\cos\varphi_i\Big)\;,\nonumber\\
&&{\cal X}_{_{U,IJK}}^{^R}=s_{_\beta}^2{\bf Re}\Big({\cal Z}_{_U}^{\dagger3I}
({\cal Z}_{_U}\hat{\bf A}_{_U}{\cal Z}_{_Q})^{IJ}
({\cal Z}_{_U}\hat{\bf A}_{_U}{\cal Z}_{_Q})^{\dagger JK}
{\cal Z}_{_U}^{\dagger K3}\Big)\;,\nonumber\\
&&{\cal X}_{_{U,IJK}}^{^A}=s_{_\beta}^2{\bf Im}\Big({\cal Z}_{_U}^{\dagger3I}
({\cal Z}_{_U}\hat{\bf A}_{_U}{\cal Z}_{_Q})^{IJ}
({\cal Z}_{_U}\hat{\bf A}_{_U}{\cal Z}_{_Q})^{\dagger JK}
{\cal Z}_{_U}^{\dagger K3}\Big)\;,\nonumber\\
&&{\cal X}_{_{q,IJK}}^{^S}=\chi_{_q}{\bf Re}\Big(({\bf h}_{_U}{\cal Z}_{_Q})^{3I}
({\cal Z}_{_q}\hat{\bf A}_{_q}{\cal Z}_{_Q})^{\dagger IJ}
({\cal Z}_{_q}{\bf A}_{_q}{\cal Z}_{_Q})^{JK}
({\cal Z}_{_Q}^\dagger{\bf h}_{_U}^\dagger)^{3I}\Big)
\;,\;(q=U,\;D;\;\chi_{_U}=s_{_\beta}^2,\;\chi_{_D}=c_{_\beta}^2)\;,\nonumber\\
&&{\cal X}_{_{q,IJK}}^{^B}=\chi_{_q}{\bf Im}\Big(({\bf h}_{_U}{\cal Z}_{_Q})^{3I}
({\cal Z}_{_q}\hat{\bf A}_{_q}{\cal Z}_{_Q})^{\dagger IJ}
({\cal Z}_{_q}\hat{\bf A}_{_q}{\cal Z}_{_Q})^{JK}
({\cal Z}_{_Q}^\dagger{\bf h}_{_U}^\dagger)^{3I}\Big)
\;,\;(q=U,\;D;\;\chi_{_U}=s_{_\beta}^2,\;\chi_{_D}=c_{_\beta}^2)\;,\nonumber\\
&&{\cal X}_{_{q,IJK}}^{^T}=\chi_{_q}{\bf Re}\Big[{\cal Z}_{_Q}^{3I}\Big({\cal Z}_{_q}\hat{\bf A}_{_q}
{\cal Z}_{_Q}\Big)^\dagger_{IJ}\Big({\cal Z}_{_q}\hat{\bf A}_{_q}{\cal Z}_{_Q}\Big)_{JK}
{\cal Z}_{_Q}^{\dagger K3}\Big]
\;,(q=U,\;D;\;\chi_{_U}=s_{_\beta}^2,\;\chi_{_D}=c_{_\beta}^2)\;,\nonumber\\
&&{\cal X}_{_{q,IJK}}^{^C}=\chi_{_q}{\bf Im}\Big[{\cal Z}_{_Q}^{3I}\Big({\cal Z}_{_q}\hat{\bf A}_{_q}
{\cal Z}_{_Q}\Big)^\dagger_{IJ}\Big({\cal Z}_{_q}\hat{\bf A}_{_q}{\cal Z}_{_Q}\Big)_{JK}
{\cal Z}_{_Q}^{\dagger K3}\Big]
\;,(q=U,\;D;\;\chi_{_U}=s_{_\beta}^2,\;\chi_{_D}=c_{_\beta}^2)\;,\nonumber\\
&&{\cal X}_{_{q,IJK}}^{^U}=\chi_{_q}{\bf Re}\Big[\Big({\bf h}_{_q}^\dagger{\cal Z}_{_q}^\dagger\Big)_{3I}
\Big({\cal Z}_{_q}\hat{\bf A}_{_q}{\cal Z}_{_Q}\Big)_{IJ}\Big({\cal Z}_{_q}\hat{\bf A}_{_q}
{\cal Z}_{_Q}\Big)^\dagger_{JK}\Big({\cal Z}_{_q}{\bf h}_{_q}\Big)_{K3}\Big]
\;,(q=U,\;D;\;\chi_{_U}=s_{_\beta}^2,\;\chi_{_D}=c_{_\beta}^2)\;,\nonumber\\
&&{\cal X}_{_{q,IJK}}^{^D}=\chi_{_q}{\bf Im}\Big[\Big({\bf h}_{_q}^\dagger{\cal Z}_{_q}^\dagger\Big)_{3I}
\Big({\cal Z}_{_q}\hat{\bf A}_{_q}{\cal Z}_{_Q}\Big)_{IJ}\Big({\cal Z}_{_q}\hat{\bf A}_{_q}
{\cal Z}_{_Q}\Big)^\dagger_{JK}\Big({\cal Z}_{_q}{\bf h}_{_q}\Big)_{K3}\Big]
\;,(q=U,\;D;\;\chi_{_U}=s_{_\beta}^2,\;\chi_{_D}=c_{_\beta}^2)\;.
\label{coup}
\end{eqnarray}



\begin{figure}
\setlength{\unitlength}{1mm}
\begin{center}
\begin{picture}(230,200)(50,90)
\put(50,80){\includegraphics{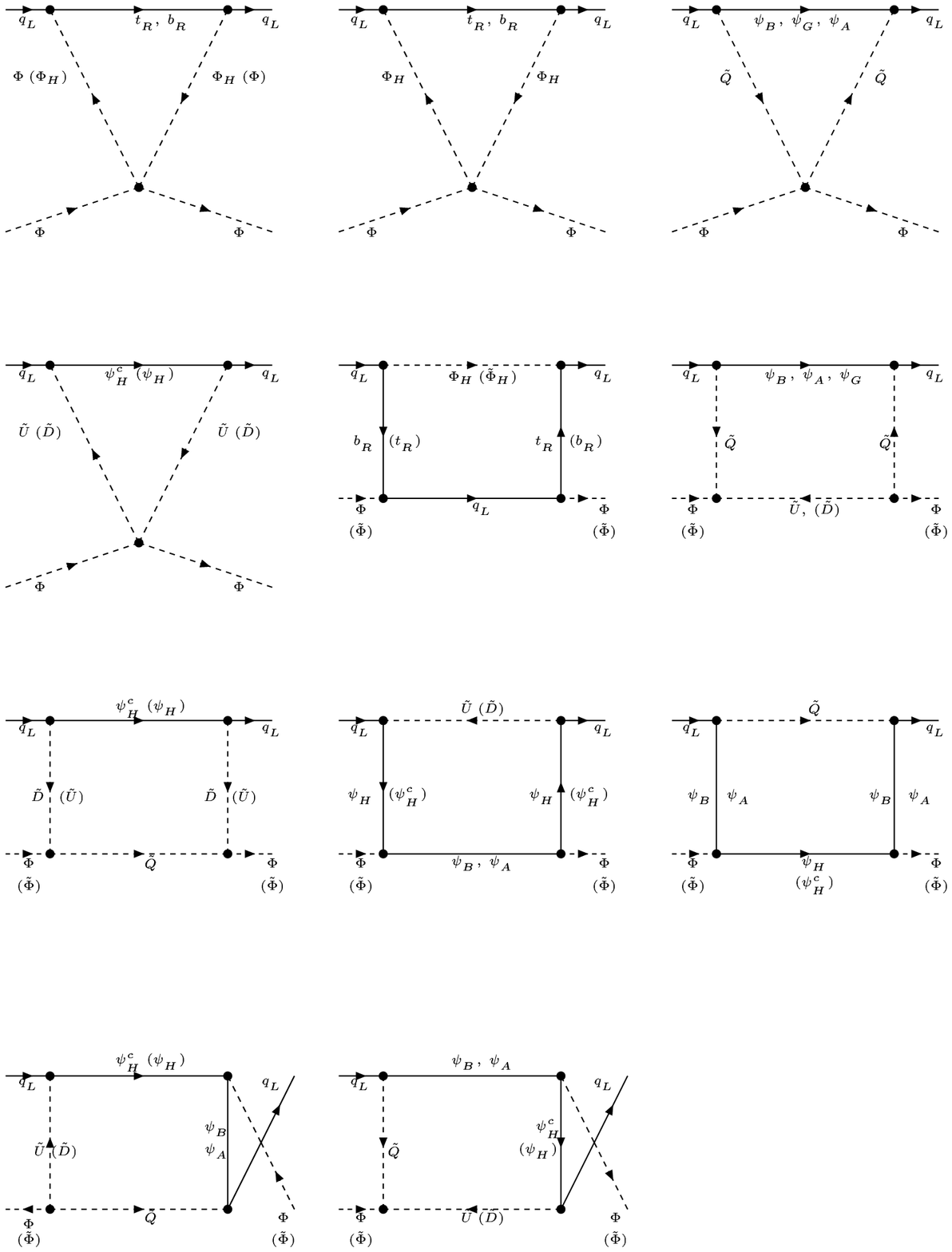}}
\end{picture}
\caption[]{The Feynman diagrams which induce  nontrivial
contributions to the Wilson coefficients of the operators ${\cal
O}_{_{q\Phi i}}\;(i= 1,\;\cdots,\;6)$.} \label{fig8}
\end{center}
\end{figure}

\begin{figure}
\setlength{\unitlength}{1mm}
\begin{center}
\begin{picture}(230,200)(55,90)
\put(50,80){\includegraphics{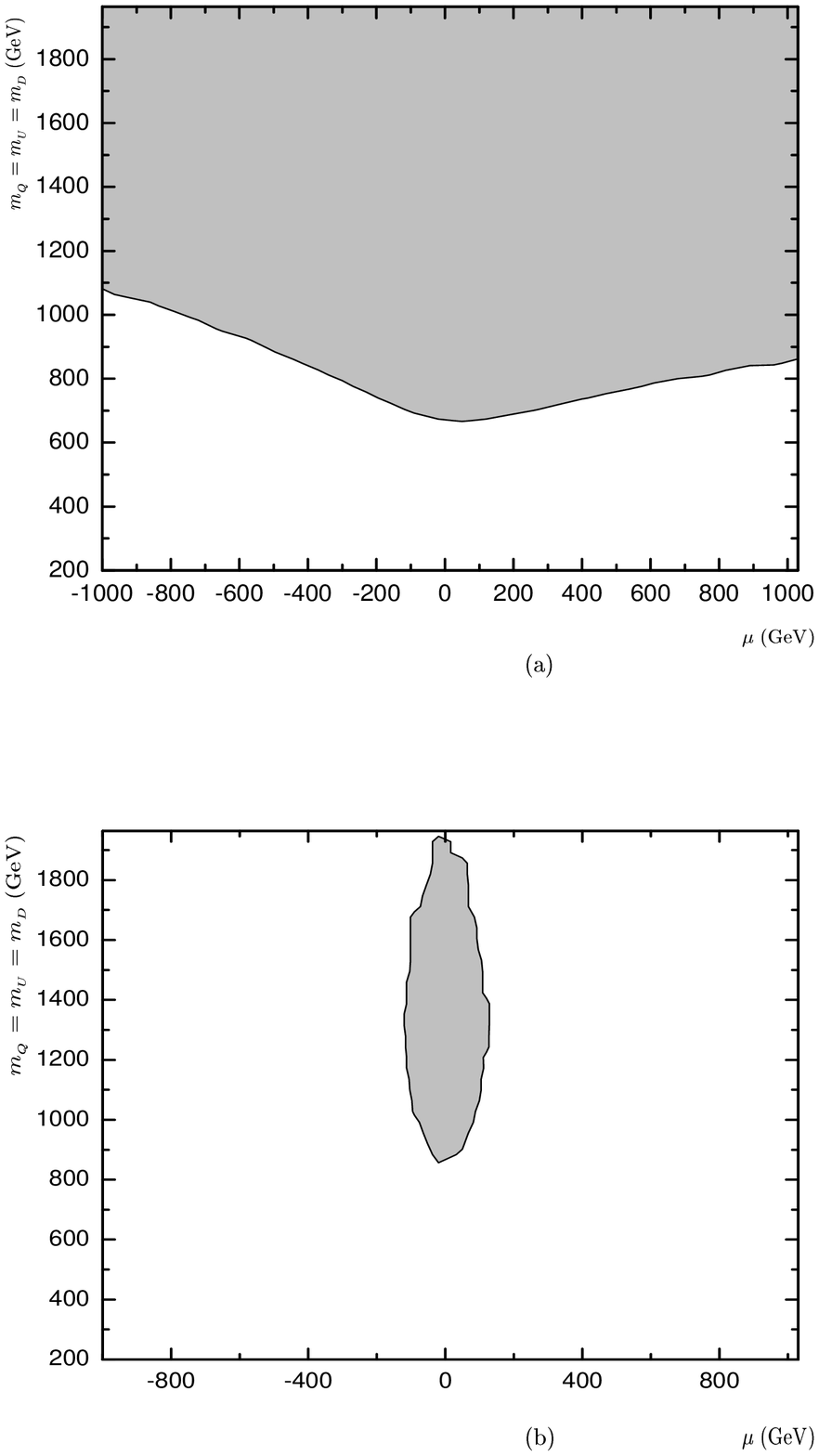}}
\end{picture}
\caption[]{The constraint from the anomalous coupling ${\cal
O}_{_{q\Phi1,2}}$ set by the $R_b$ experimental data with  $1\sigma$
tolerance,on the soft breaking parameters $m_{_Q}=m_{_U}=m_{_D}$
versus the $\mu$ parameter in the superpotential with
$m_1=m_2=m_3=500\; {\rm GeV},\;A_t=A_b=100 \;{\rm GeV},
\;m_{_H}=500\; {\rm GeV}$ and (a) $\tan\beta=2$; (b)
$\tan\beta=40$.} \label{fig9}
\end{center}
\end{figure}
\begin{figure}
\setlength{\unitlength}{1mm}
\begin{center}
\begin{picture}(230,200)(55,90)
\put(50,80){\includegraphics{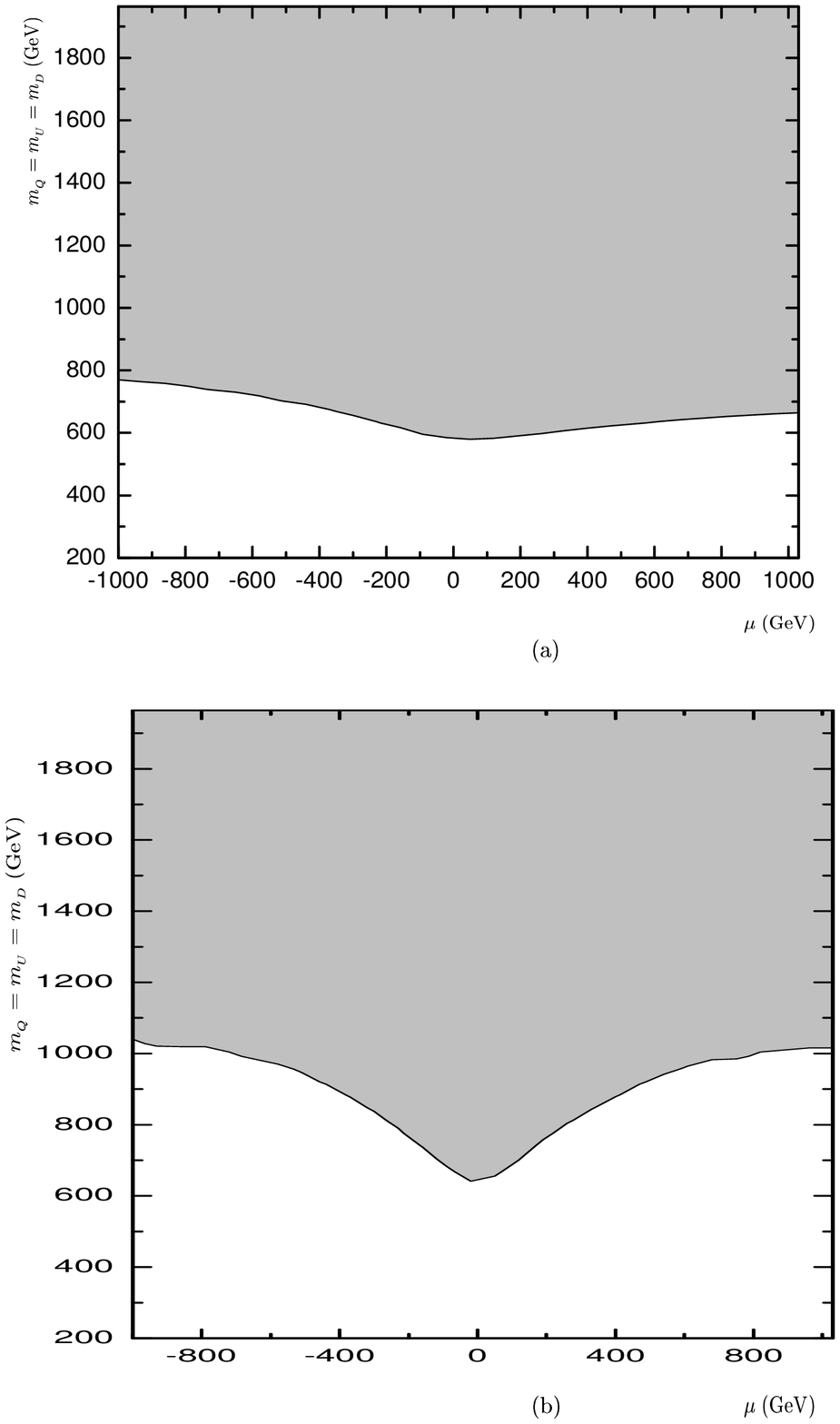}}
\end{picture}
\caption[]{Relaxing the lower bound to $-5\times10^{-5}$ and
keeping the upper bound unchanged as in Eq.(\ref{zbb6}), the
constraint from the anomalous coupling ${\cal O}_{_{q\Phi1,2}}$ on
the soft breaking parameters $m_{_Q}=m_{_U}=m_{_D}$ versus the
$\mu$ parameter in the superpotential with $m_1=m_2=m_3=500\; {\rm
GeV},\;A_t=A_b=100\; {\rm GeV}, \;m_{_H}=500\; {\rm GeV}$ and (a)
$\tan\beta=2$; (b) $\tan\beta=40$.}\label{fig10}
\end{center}
\end{figure}
\begin{figure}
\setlength{\unitlength}{1mm}
\begin{center}
\begin{picture}(230,200)(55,90)
\put(50,80){\includegraphics{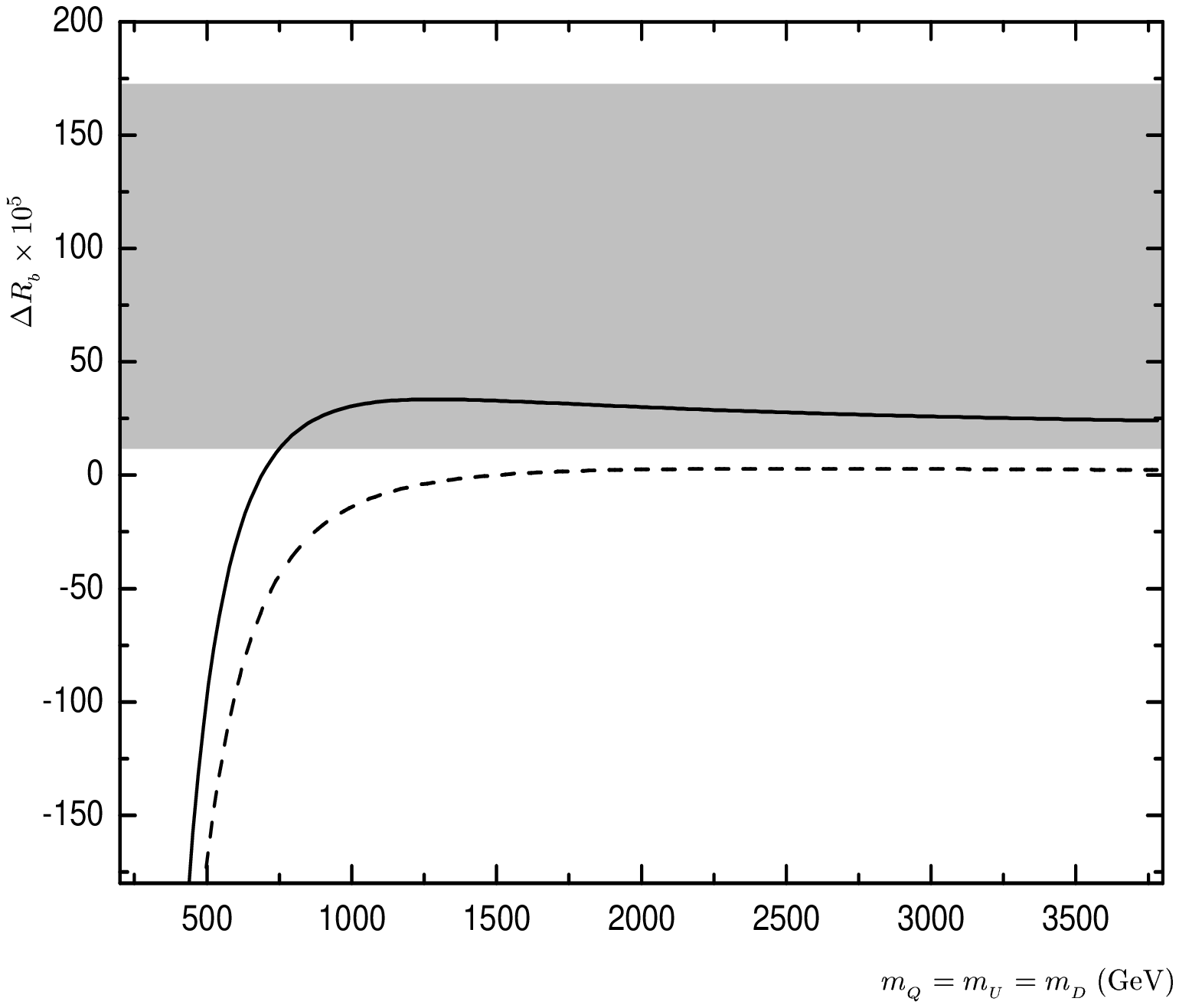}}
\end{picture}
\caption[]{Taking $\mu=m_i=500\;{\rm GeV}\;(i=1,\;2,\;3)
,\;A_t=A_b=100\; {\rm GeV}, \;m_{_H}=500\; {\rm GeV}$, $\Delta
R_{_b}$ versus squark masses $m_{_Q}=m_{_U}=m_{_D}$ with
$\tan\beta=2$ (solid line) or $\tan\beta=40$ (dashed line).
}\label{fig11}
\end{center}
\end{figure}
\begin{figure}
\setlength{\unitlength}{1mm}
\begin{center}
\begin{picture}(230,200)(55,90)
\put(50,80){\includegraphics{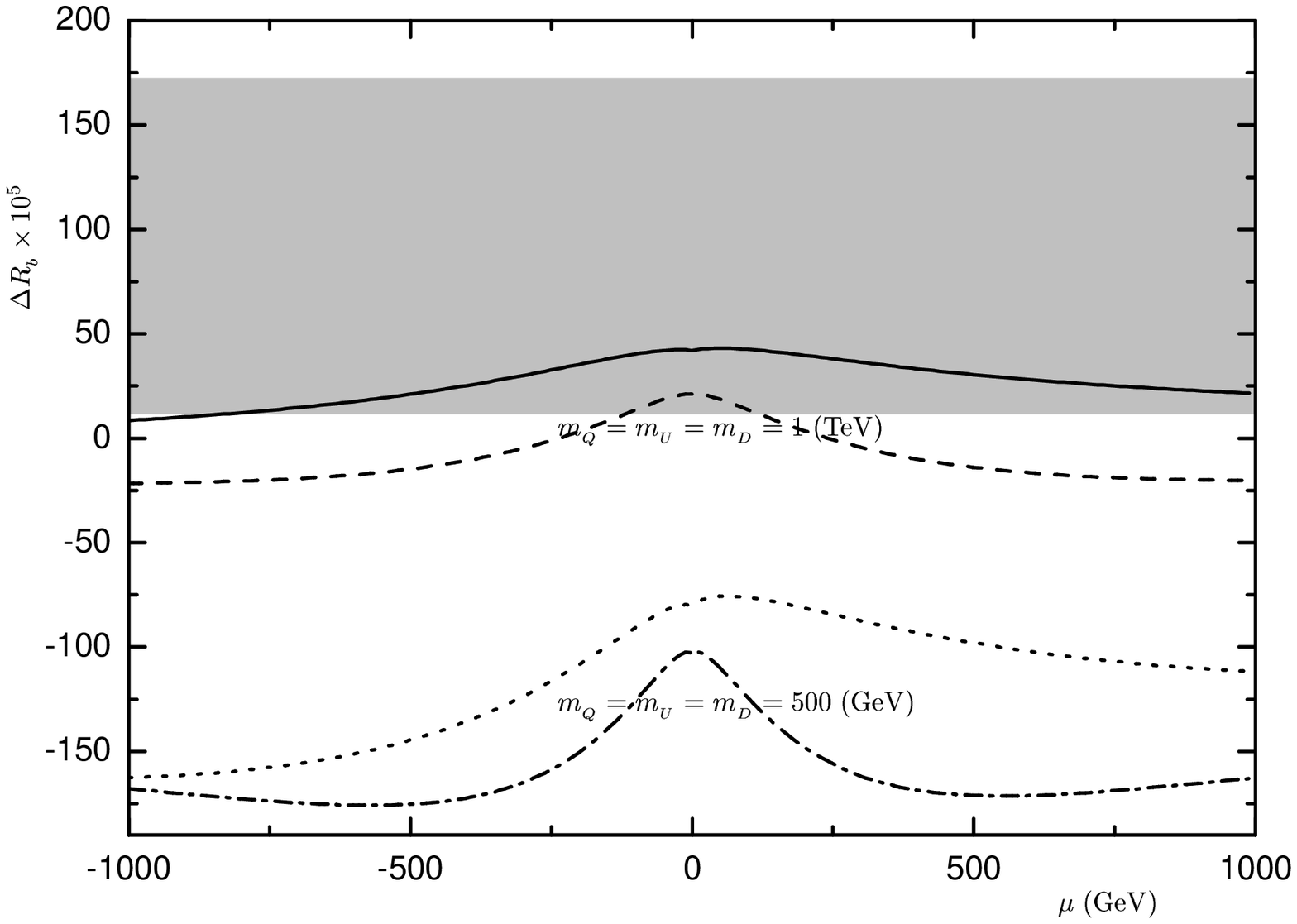}}
\end{picture}
\caption[]{$\Delta R_{_b}$ versus the parameter $\mu$ with
$m_i=500\;{\rm GeV}\;(i=1,\;2,\;3) ,\;A_t=A_b=100\; {\rm GeV},
\;m_{_H}=500\; {\rm GeV}$, and (a) solid-line:
$m_{_Q}=m_{_U}=m_{_D}=1\;{\rm TeV},\;\tan\beta=2$, (b)
dashed-line: $m_{_Q}=m_{_U}=m_{_D}=1\;{\rm TeV},\;\tan\beta=40$,
(c) dot-line: $m_{_Q}=m_{_U}=m_{_D}=500\;{\rm GeV},\;\tan\beta=2$,
(d) dot-dashed-line: $m_{_Q}=m_{_U}=m_{_D}=500\;{\rm
GeV},\;\tan\beta=40$. }\label{fig12}
\end{center}
\end{figure}
\begin{figure}
\setlength{\unitlength}{1mm}
\begin{center}
\begin{picture}(230,200)(55,90)
\put(50,80){\includegraphics{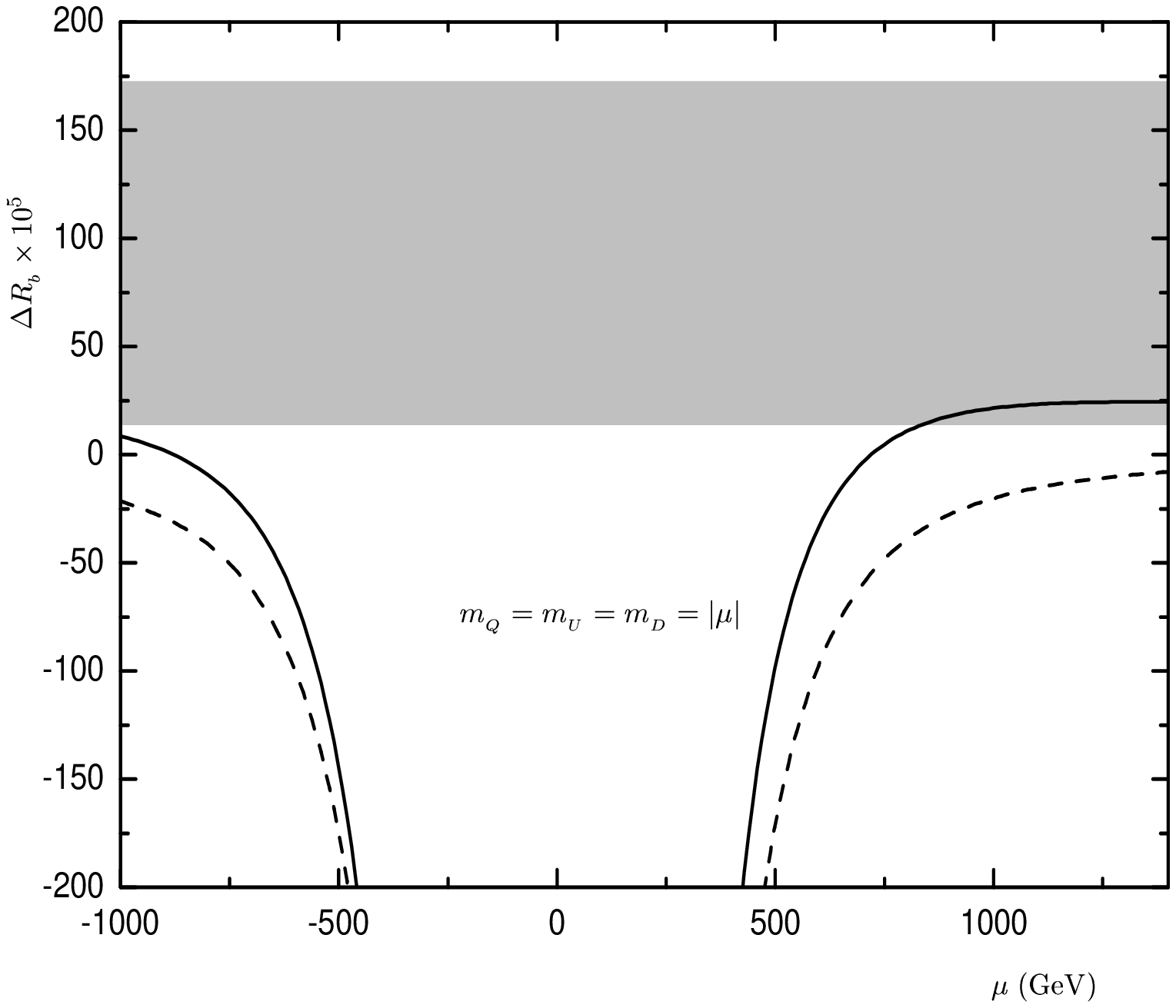}}
\end{picture}
\caption[]{Under the assumption $m_{_Q}=m_{_U}=m_{_D}=|\mu|$,
$\Delta R_{_b}$ versus the parameter $\mu$. The other parameters
are taken as $m_i=500\;{\rm GeV}\;(i=1,\;2,\;3) ,\;A_t=A_b=100\;
{\rm GeV}, \;m_{_H}=500\; {\rm GeV}$ and $\tan\beta=2$
(solid-line) $\tan\beta=40$ (dashed-line). }\label{fig13}
\end{center}
\end{figure}
\end{document}